\documentclass[11pt]{article}
\pdfoutput=1 % if your are submitting a pdflatex (i.e. if you have
\usepackage{jheppub} % for details on the use of the package, please
\usepackage[T1]{fontenc} % if needed
\usepackage{amssymb,amsbsy,bm,amsmath,psfrag,dsfont}
\usepackage{graphicx}
\usepackage{epsfig}
\usepackage{subfigure}
\usepackage{float}

\makeatletter
\def\@fpheader{\relax}
\makeatother

\title{\boldmath On the genetic architecture of intelligence and other quantitative traits}

\author[]{Stephen D.H. Hsu}
\affiliation[]{Department of Physics and Astronomy, Michigan State University \\ Cognitive Genomics Laboratory, BGI \\
{\rm hsu@msu.edu}}
\abstract{How do genes affect cognitive ability or other human quantitative traits such as height or disease risk? Progress on this challenging question is likely to be significant in the near future. I begin with a brief review of psychometric measurements of intelligence, introducing the idea of a "general factor" or g score. The main results concern the stability, validity (predictive power), and heritability of adult g. The largest component of genetic variance for both height and intelligence is additive (linear), leading to important simplifications in predictive modeling and statistical estimation. Due mainly to the rapidly decreasing cost of genotyping, it is possible that within the coming decade researchers will identify loci which account for a significant fraction of total g variation. In the case of height analogous efforts are well under way. I describe some unpublished results concerning the genetic architecture of height and cognitive ability, which suggest that roughly 10k moderately rare causal variants of mostly negative effect are responsible for normal population variation. Using results from Compressed Sensing (L1-penalized regression), I estimate the statistical power required to characterize both linear and nonlinear models for quantitative traits. The main unknown parameter $s$ (sparsity) is the number of loci which account for the bulk of the genetic variation. The required sample size is of order $100 \, s$, or roughly a million in the case of cognitive ability. 

\bigskip
\noindent These informal notes are based on lectures presented at a number of universities and research institutes from 2011-2014, including BGI, the University of Oregon, UC Irvine, UC Davis, Academia Sinica (Taiwan), National Center for Theoretical Physics and Mathematics (Taiwan), BGI Shenzhen, Google, Caltech, Michigan State University, and the University of Chicago.}

\begin{document} 
\maketitle
\flushbottom

\begin{quote}
\noindent {\it You know Herb, how much faster I am in thinking than you are. That is how much faster von Neumann is compared to me.} -- Nobel Laureate Enrico Fermi to his former doctoral student, University of Chicago Physics Professor Herb Anderson.
\smallskip

\noindent {\it I always thought von Neumann's brain indicated that he was from another species, an evolution beyond man.} -- Nobel Laureate Hans A. Bethe.
\end{quote}
\bigskip

\newpage

\section{Introduction: Genomics in the 21st century}
%\label{sec:intro}

The human brain is perhaps the most complex object we know of in the universe. Yet, it is constructed from genetic instructions of modest size -- roughly a few gigabits of information. Decoding how Nature builds the brain is one of the most challenging of scientific mysteries.

\begin{figure}[tbph]
\begin{center}
\includegraphics[width=12cm]{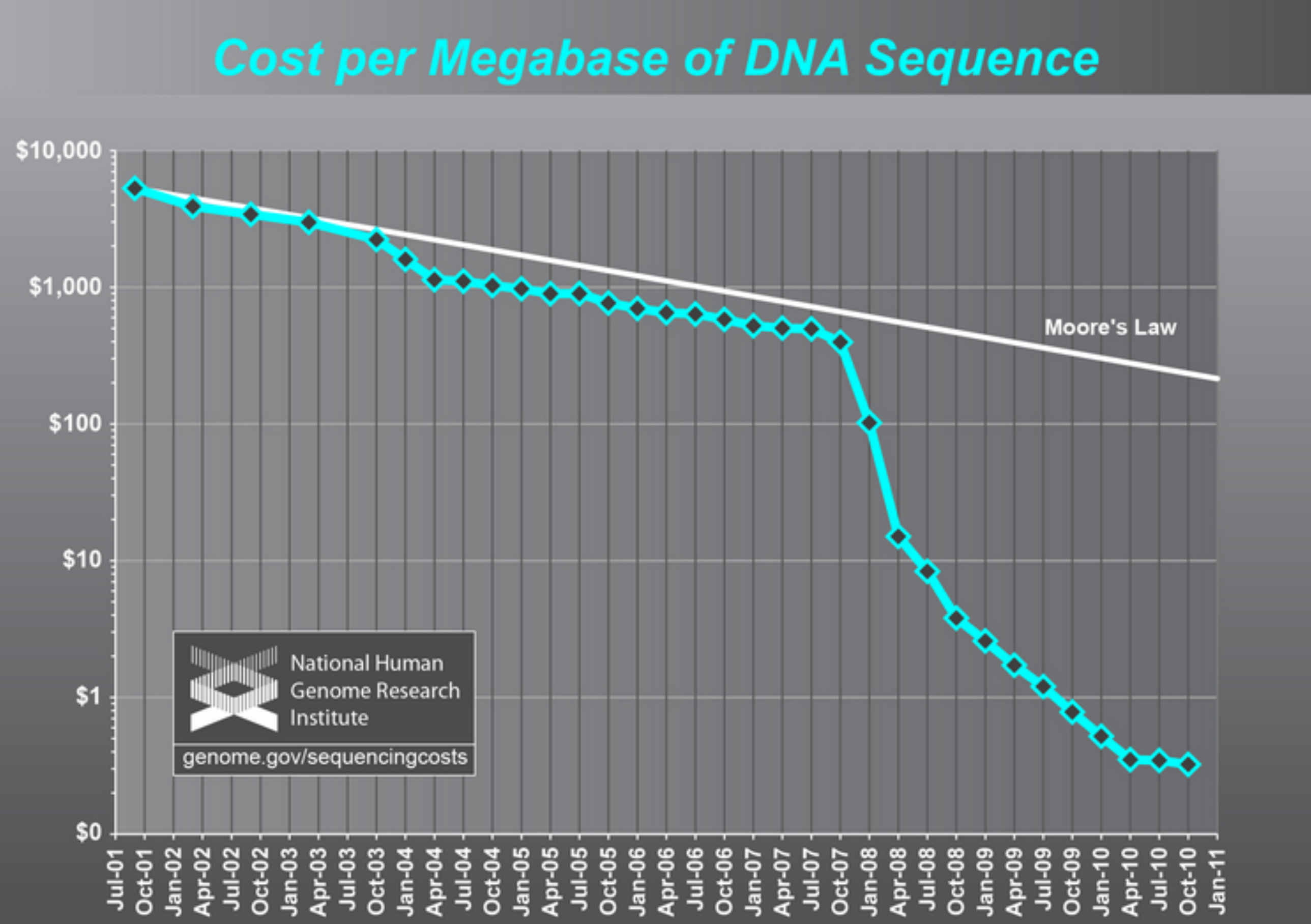}
\end{center}
\caption{Sequencing costs are declining at a super-exponential rate.}
\label{costs}
\end{figure}

Technical advances now allow us to read the genetic code of an individual organism. The cost to do so has decreased recently at a super-exponential rate (Fig.(\ref{costs})), thanks to a confluence of factors: government investment in basic research, venture capital and public market investment in risky new technologies, and an influx of human capital into genomics from engineering and the physical and information sciences. Large data sets of human genotypes and phenotypes will lead to significant progress in our ability to understand the genetic code -- in particular, to predict phenotype from genotype.

\begin{figure}[tbph]
\begin{center}
\includegraphics[width=12cm]{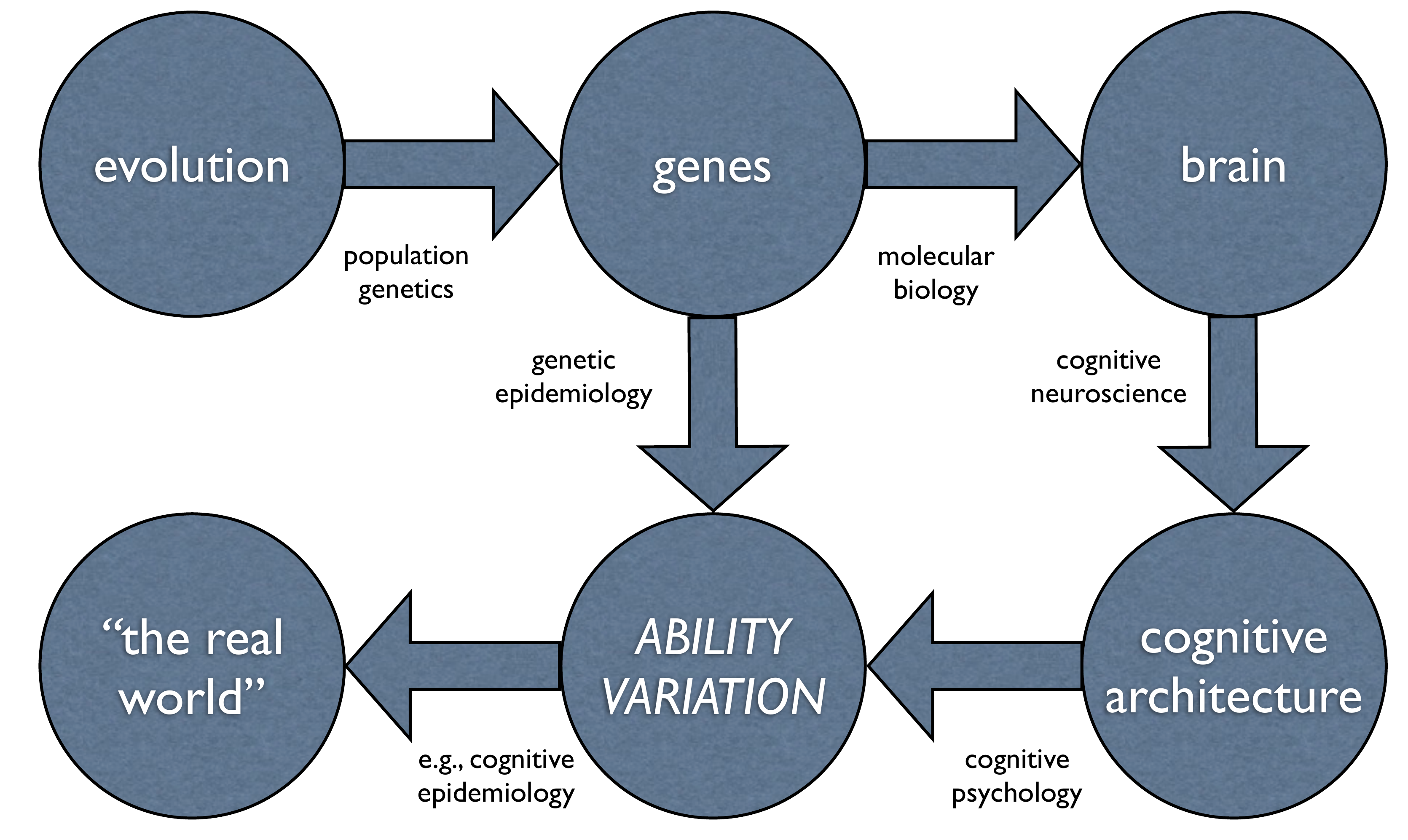}
\end{center}
\caption{Disciplines impacted by new genomic technologies.}
\label{disciplines}
\end{figure}

Quantitative traits are highly polygenic: influenced by many genes of small effect. Some geneticists are skeptical that we will ever understand traits that are influenced by hundreds or even thousands of genes. However, as I argue below, even phenotypes with complex genetic architecture will yield to genomic modeling once a sufficient amount of genotype$\vert$phenotype data becomes available. In the case of intelligence and height, I predict this will happen within the next 10 years. To be clear, what I mean here is {\it not} that we will grasp all of the secrets of the operation or construction of the brain. Rather, and much more modestly, we will be able to {\it predict} cognitive ability level (as defined in the next section) from genotype. Nevertheless, progress on this more limited problem will inform many areas of research related to the brain and cognition (Fig.(\ref{disciplines})): identification of loci whose variation influences cognition will implicate specific proteins, structures, neurotransmitters, chemical pathways and mechanisms in the brain. We will also gain insights into human evolution and how natural selection led to human intelligence. Our discussion mainly addresses the genes $\rightarrow$ ability variation part of the diagram in Fig.(\ref{disciplines}), nevertheless it is clear that the results will impact, to some degree, all of the other regions.

It is worthwhile to summarize some rough numbers characterizing human genetics. The genome consists of $\sim 3 \times 10^9$ base pairs, with inter-individual variation at about 1 in 1000 loci, leading to a few million differences on average between two individuals. Due to this limited variation, genomes in bulk are compressible to few megabytes of information, specifying individual deviation from a reference sequence. In comparison, humans differ from chimpanzees at roughly 1 per 100 loci, and from Neanderthals at roughly few per 1000 loci. It is interesting that significant differences in physical morphology and cognitive ability can result from relatively few genetic modifications.

Regions exhibiting common variation (i.e., where two humans have probability of at least a few percent of differing; there are roughly $10^7$ such sites) can be tagged by SNPs, or Single Nucleotide Polymorphisms. SNP chips allow an inexpensive but informative sampling of the whole genome. Whole genome sequencing will allow access to additional forms of structural variation not well tagged by SNPs. In 2014, the cost of SNP genotyping is $\sim \$100$ USD, and for whole genome sequencing $\sim \$1000$. Sample sizes for SNP genotypes are well into the hundreds of thousands. If all available SNP genotypes were aggregated the total number might exceed one million. Note, however, that institutional limitations are an obstacle to this aggregation of data. It is also likely that no specific phenotype is simultaneously available for all of these genotypes.

These notes are deliberately brief and informal. They are intended for a multidisciplinary audience: progress in this area of research will require expertise from a broad range of specialties, including population and quantitative genetics, sequencing and genotyping technologies, molecular biology, psychology, cognitive science, neuroscience, algorithms and computation, statistics and applied mathematics. For more information and additional references, see the BGI Cognitive Genomics project document \url{https://www.cog-genomics.org/static/pdf/bgi_g_proposal.pdf} and FAQ: \url{https://www.cog-genomics.org/faq}. Some items from the FAQ are reproduced at the end of this paper.

\section{Cognitive ability and heritability}
%\label{sec:cog}

\begin{figure}[tbph]
\begin{center}
\includegraphics[width=12cm]{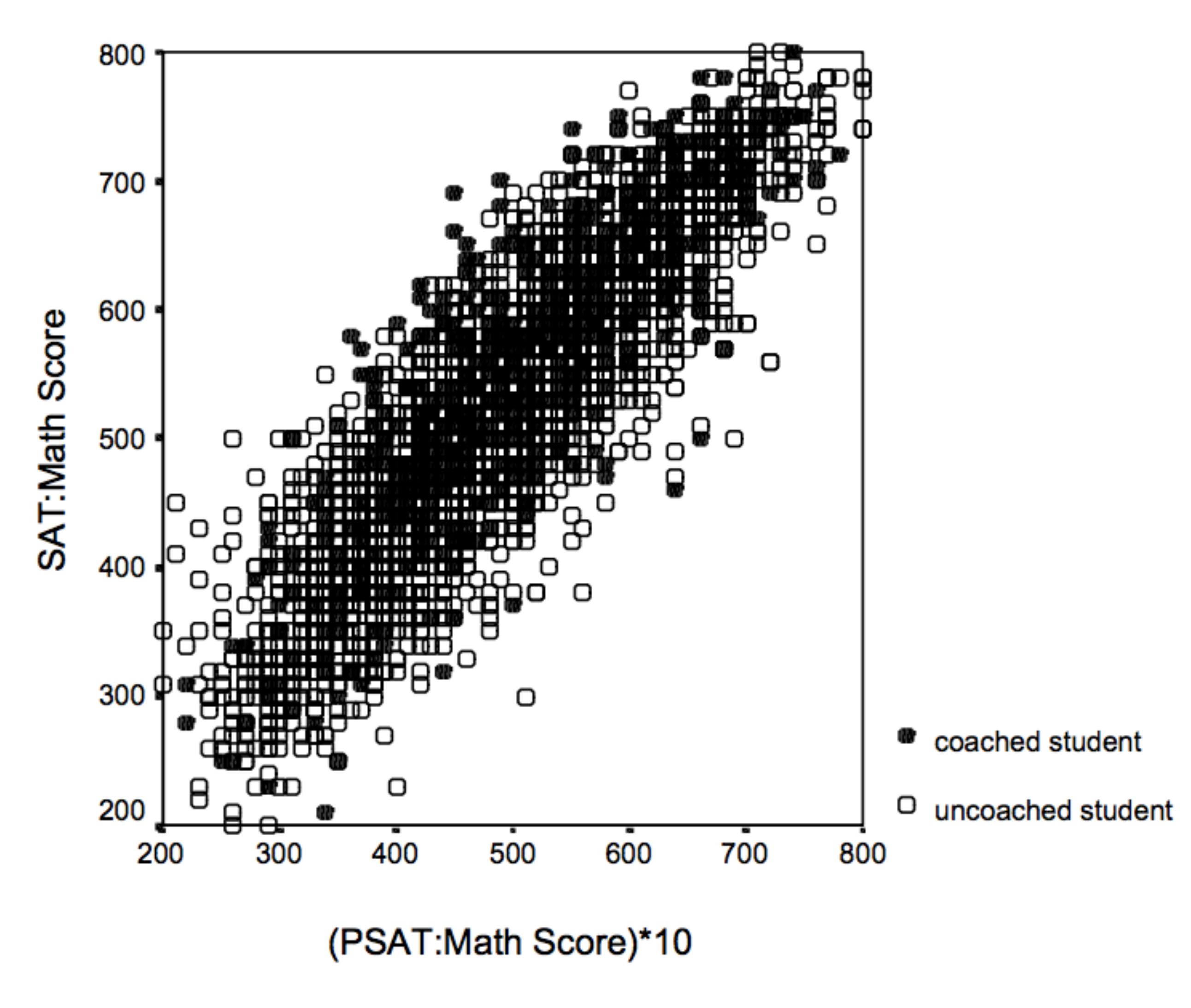}
\end{center}
\caption{Test preparation has little impact on SAT scores.}
\label{fig:testprep}
\end{figure}

While perhaps unfamiliar to non-specialists, the quantitative study of cognitive ability (psychometrics) has produced a large body of robust results \cite{greview}. Standardized testing is in widespread use in pre-university (PISA, TIMSS), university (SAT, ACT, GRE) and military (AFQT) contexts. Many advanced countries use cognitive testing as a means to allocate scarce educational resources and to predict which individuals are likely to succeed in both ordinary and demanding job categories. Cognitive ability can be measured, at least crudely, and constitutes a {\it quantitative phenotype} suitable for genomic investigation. The main qualities required are:

 \bigskip
1. Stability, Reliability: Test-retest correlations are typically in the 0.9 to 0.95 range, even for administrations a year apart \cite{greview}. Effect of commercial test preparation on standardized scores (SAT, ACT) is much less than a population standard deviation (SD) (Fig.(\ref{fig:testprep}) and \cite{testprep}). Rank order of scores (taking into account measurement error) tends to be stable over adult lifetime.

\smallskip
2. Validity: cognitive scores are predictive of life outcomes, job performance, academic performance, even longevity after controlling for social status \cite{greview}. For example, correlation with university performance is $\sim 0.4$, and minimum thresholds are observed for mastery of certain subjects (e.g., physics or mathematics) \cite{HsuSchombert}.

\smallskip
3. Heritability (genetic causes): under good environmental conditions, a large portion (exceeding half) of the variation in cognitive ability is probably due to genetic variation \cite{heritability}. See below for more discussion.
\bigskip

Our ability to measure cognitive ability is crude, and it is easy to point to limitations in existing methodology. However, even a flawed measure is useful if it is positively correlated with intelligence. It is plausible that, on average, individuals with higher cognitive scores have better functioning brains -- at least, brains which are better at processing information and dealing with abstraction.

Below we discuss how one arrives at a measure referred to as the general factor of cognitive ability, or ``g''. Once defined, the quantity should seem quite reasonable, leaving only the empirical question of whether it satisfies properties 1-3 above. Perhaps surprisingly, g scores are roughly comparable to height on each of the above criteria.

\subsection{The general factor, g}

Clearly, certain specialized abilities play a role in cognition: short and long term memory, the use of language, the use of quantities and numbers, the visualization of geometric relationships, pattern recognition, etc. A priori, one does not know whether these are entirely independent capabilities, or correlated in some way. An important observation in psychometrics (by now well-established from literally millions of observations) is that essentially every ``primitive'' cognitive ability of the type described above is {\it positively correlated}. That is, an individual who is above average in one area (e.g., mathematical ability) is more likely to be above average in another (e.g., verbal ability). Note that this result is {\it non-obvious}: it is a common folk view that some of these abilities are anti-correlated (``Johnny is good at math, therefore not so good at words''). In fact, the conditional probability that Johnny is good at words is increased by the fact that he is good at math.  

\begin{figure}[tbph]
\begin{center}
\includegraphics[width=12cm]{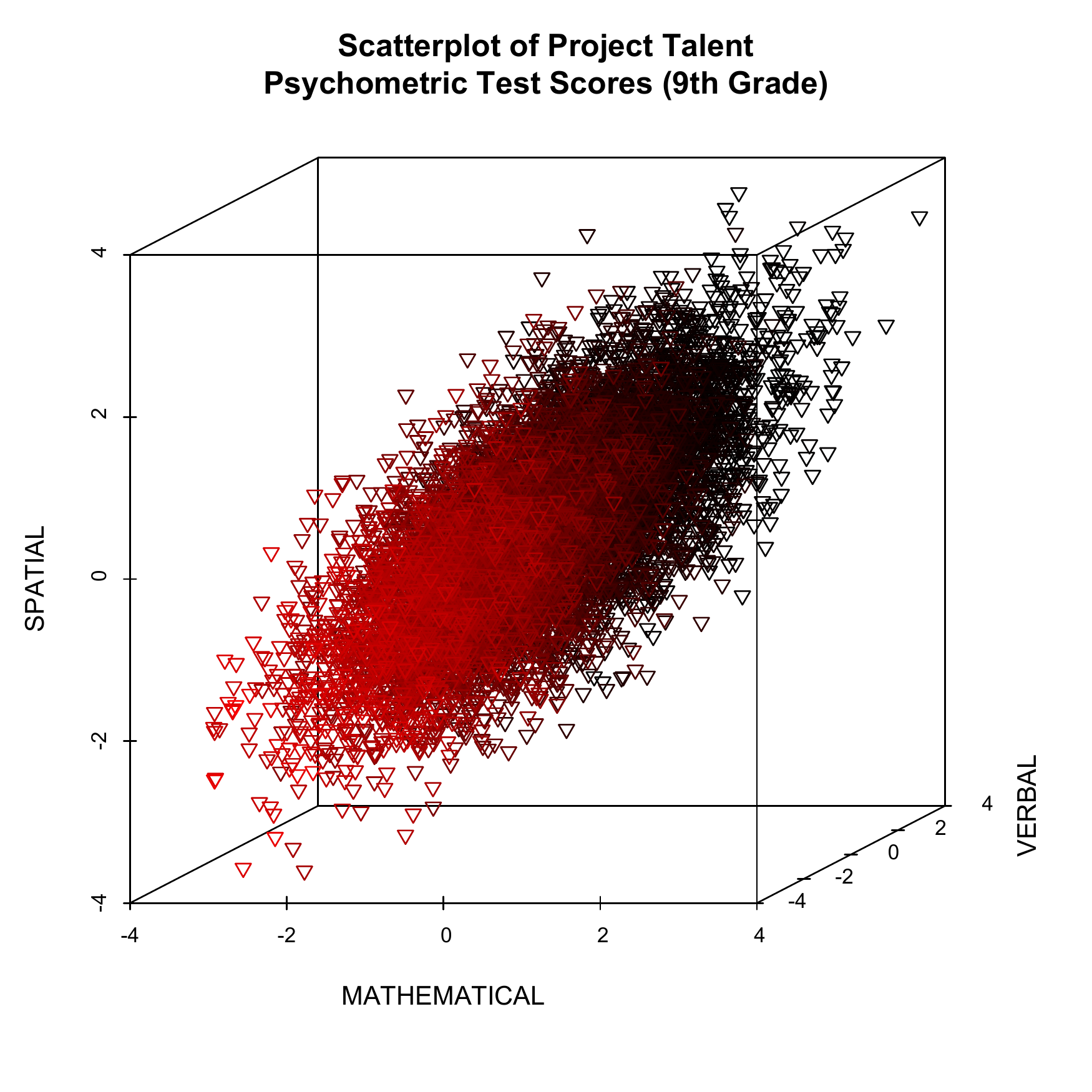}
\end{center}
\caption{Positive correlations between narrow cognitive abilities allow compressed representation of cognitive profiles. This figure is a visualization of correlations found in Project Talent data, but the individual points do not represent specific individuals.}
\label{fig:ellipsoid}
\end{figure}

These positive correlations between narrow abilities suggest a robust and useful method for {\it compressing} information concerning cognitive abilities.  I. {\it Choose a battery of n ``cognitive'' tests}, e.g.,  digit recall (short term memory), vocabulary, math puzzles, spatial rotations, reading comprehension,  $\cdots$  II. {\it Test many individuals.} This provides a map from individual to n vector. (In cases where individual ability scores are approximately normally distributed, we can imagine each entry of the n vector as a z score.) Due to the correlations, a population of individuals does not fill out the n dimensional space uniformly. Rather, they fill out an ellipsoidal region (Fig.(\ref{fig:ellipsoid})). A simple way to compress the information in the distribution is to find the principal component of variation (major axis of the ellipsoid) and project each n vector onto this axis. This leads to a single number measure of cognitive ability, the general factor g, which is (at least information-theoretically) the natural compression of the ability data \cite{psychom}. Since this definition is not unreasonable, the remaining questions are empirical: are the properties of Stability, Validity and Heritability satisfied?

\begin{figure}[tbph]
\begin{center}
\includegraphics[width=12cm]{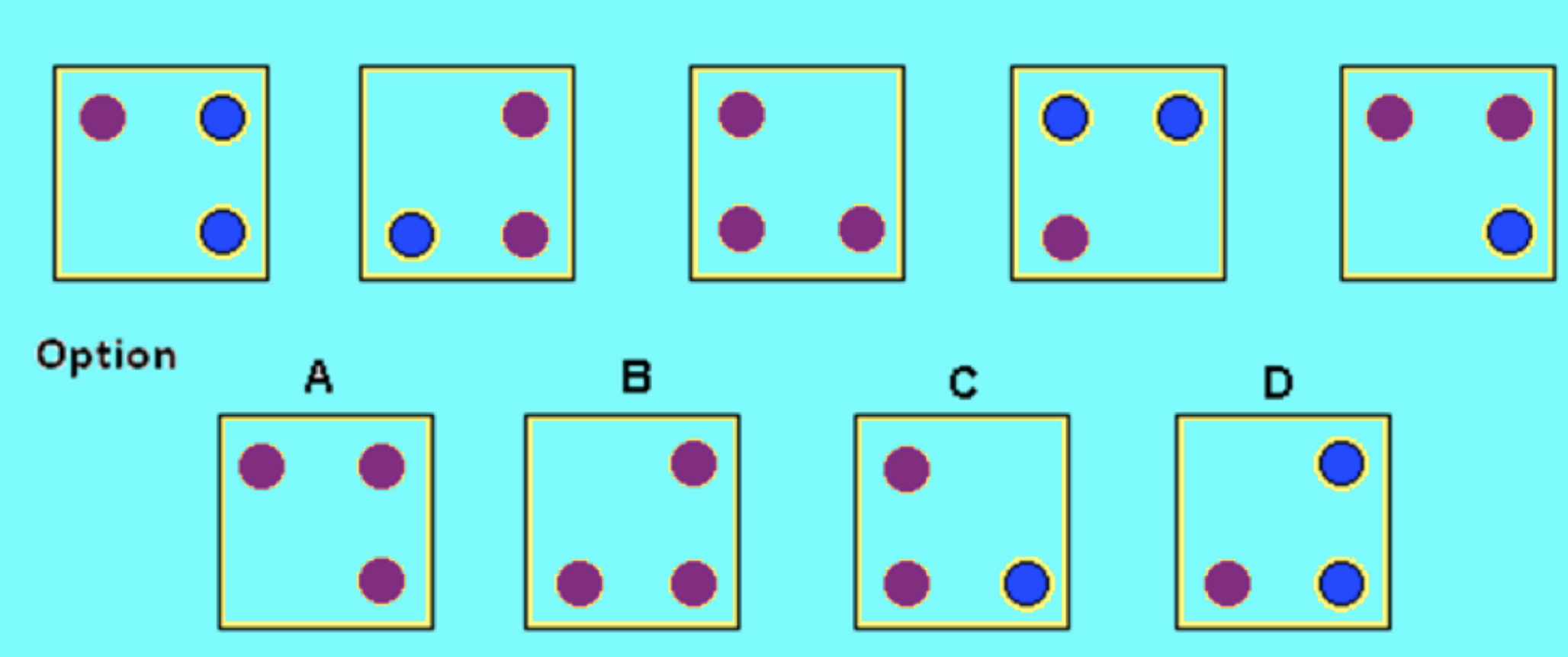}
\end{center}
\caption{Scores on abstract matrix tests are highly correlated with g. Such tests are arguably free of cultural loading and do not even require literacy.}
\label{fig:RPM}
\end{figure}

Note that the precise definition of g depends on the choice of battery of tests. There is no unique definition, but various reasonable definitions in use all correlate at $\sim (0.7 - 0.8)$. Any test that correlates at this level with g can reasonably be called a test of intelligence or cognitive ability. Interestingly, some abstract and ostensibly culturally neutral tests such as Raven's Progressive Matrices (Fig.(\ref{fig:RPM})), which tests pattern recognition and algorithmic thinking, have very high g loadings (correlation with broad battery definitions of g and low loadings on more test-specific factors). Hence, they can be used as a shortcut to obtain efficient measurements.

The goal here is not to find a perfect measure of cognitive ability (whatever that is), but rather to {\it operationalize} the measurement of a quantity that we are confident is at least {\it positively correlated} with overall intellectual ability or brain function. As long as a positive (though imperfect) correlation exists, statistical data from testing can be used to identify genomic factors that influence cognition.

\subsection{Heritability estimates}

\begin{figure}[tbph]
\begin{center}
\includegraphics[width=10cm]{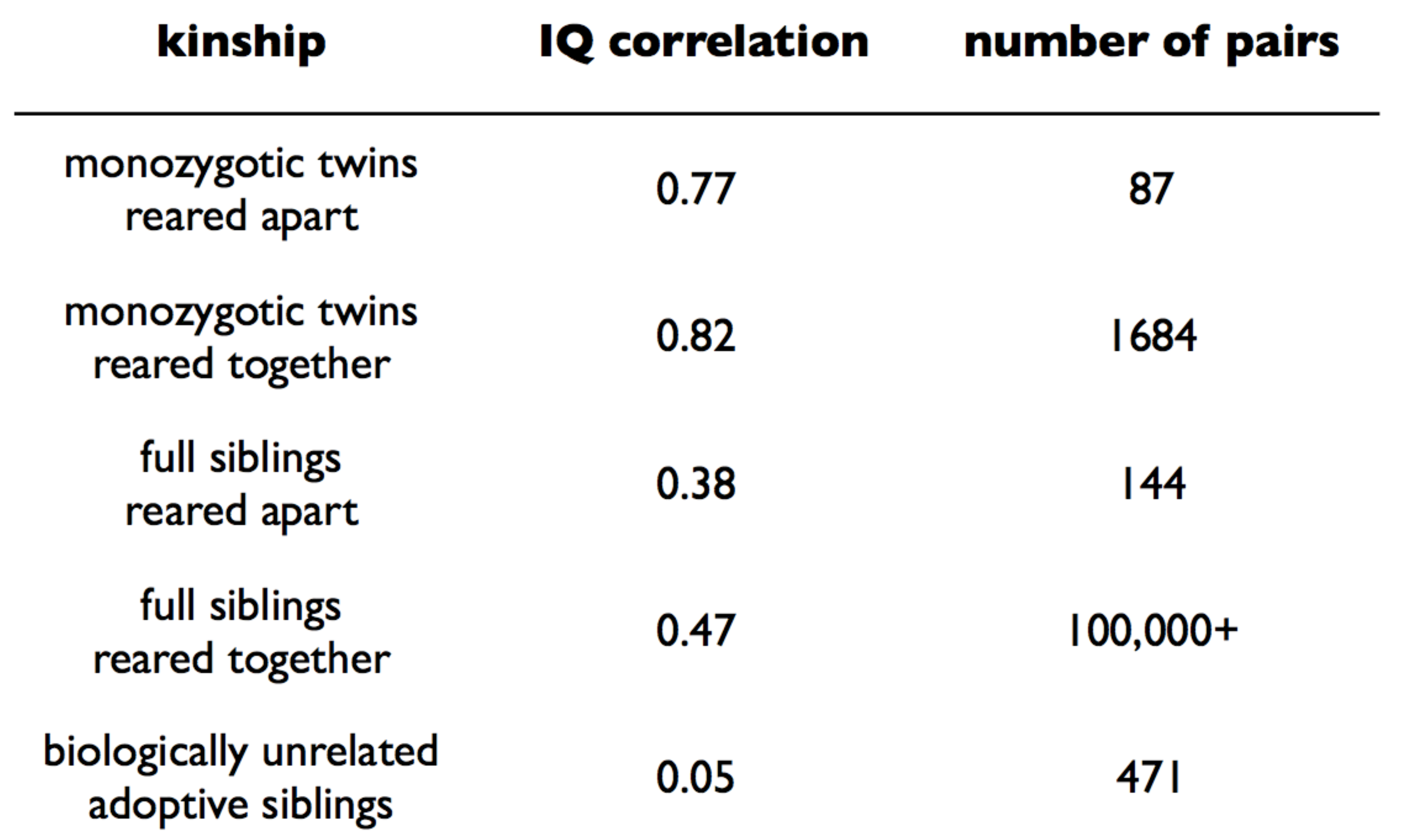}
\end{center}
\caption{IQ correlation varies approximately linearly with relatedness (fraction of genes shared). Family environment has only small impact on IQ.}
\label{twins1}
\end{figure}

\begin{figure}[tbph]
\begin{center}
\includegraphics[width=10cm]{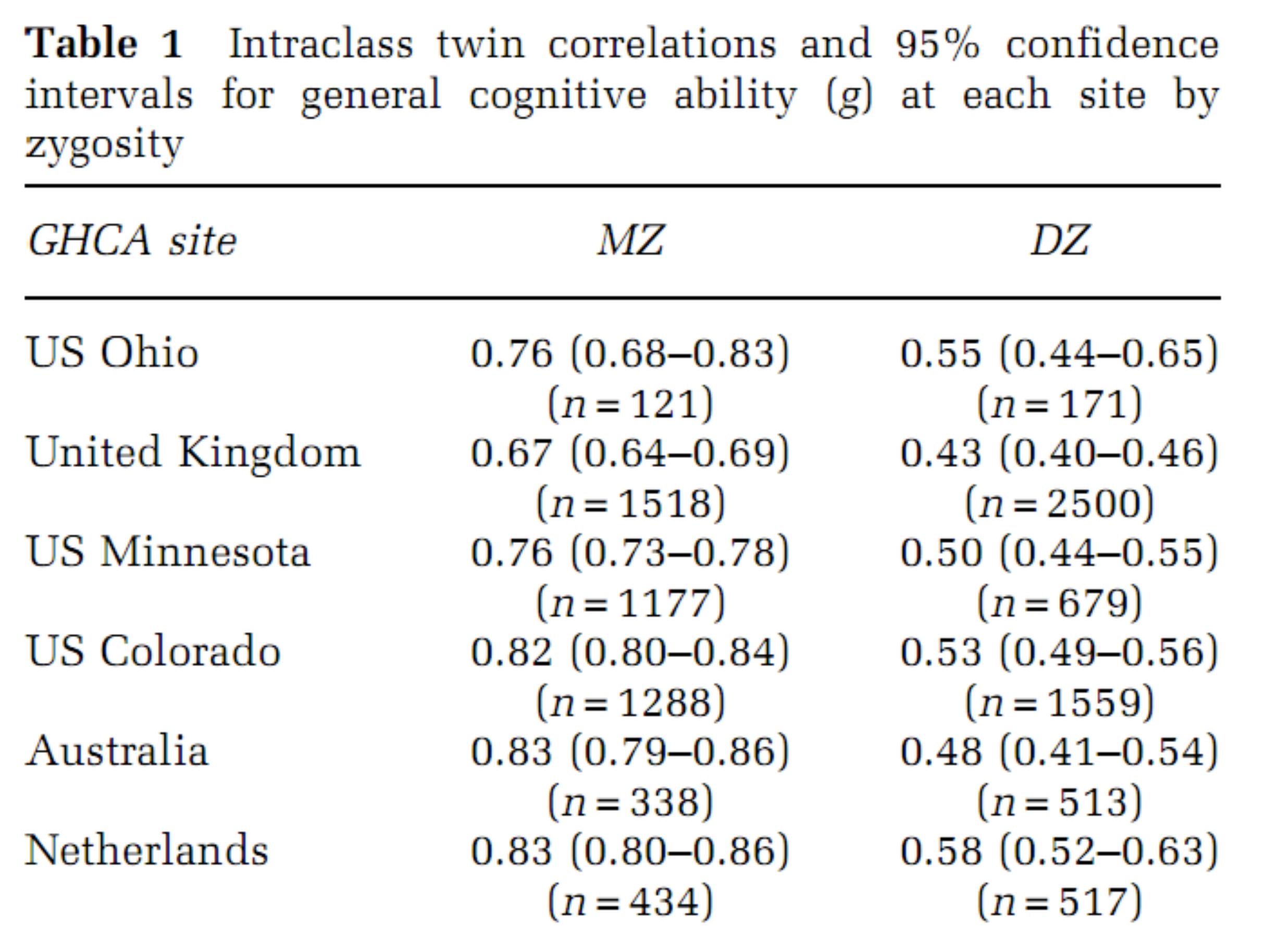}
\end{center}
\caption{Large statistics studies indicate g is highly heritable. Results are consistent across studies conducted in many countries.}
\label{twins2}
\end{figure}

The following is from Thomas Bouchard \cite{heritability}, University of Minnesota twins researcher, former President of the Behavior Genetics Association, and founder of the Minnesota Study of Twins Reared Apart: 
\begin{quote}
I will make the case for a strong genetic influence on human intelligence. I am not alone
in this claim. According to Plomin (2003, p. 108), `The case for substantial genetic
influence on g is stronger than for any other (Mackintosh 1998) human characteristic.' One
might argue that stature is more heritable, but for adult stature the findings are surprisingly
similar. For example in large twin studies conducted in Minnesota and Finland the broad
heritability of stature was 0.75 and 0.79 for men, respectively and 0.72 and 0.77 for women,
respectively (Silventoinen et al. 2004). As I will show, the results for adult intelligence are in
the same range.
\end{quote}
Fig.(\ref{twins1}) and Fig.(\ref{twins2}) give examples of heritability estimates from twin and adoption studies. In Fig.(\ref{twins1}), correlations are given for various pairs of related individuals (monozygotic (MZ) and dizygotic (DZ) twins, siblings) raised in the same family, or adopted into different families. Note that correlations are roughly proportional to degree of kinship (fraction of genes shared between the two individuals), with only small differences associated with family environment (raised together or apart). Biologically unrelated siblings raised in the same family have almost zero correlation in cognitive ability. Fig.(\ref{twins2}) gives some idea as to the consistency of these results over large studies conducted in a variety of locations.

High heritability estimates are obtained in cases where subjects have generally experienced good environments. In the absence of deprivation, it would seem that genetic effects determine the upper limit to height, cognitive ability, etc. However, in studies where subjects have experienced a wider range of environmental conditions, such as poverty, malnutrition or lack of education, heritability estimates can be much smaller \cite{Turkheimer}. When environmental conditions are unfavorable, individuals do not achieve their full potential. We discuss this further in the context of the Flynn effect below.

Genomic technologies now allow us to estimate heritability based on the correlation between phenotype similarity and genetic relatedness measured directly from DNA. These estimates use large samples of unrelated individuals and obtain results which are consistent with the ``classical'' twin and adoption designs \cite{GCTA}.

\subsection{Flynn effect}

The Flynn effect refers to a significant increase in raw cognitive scores over the last 100 years or so -- equivalent of 30 points (2 SD) in some cases. This raises a number of thorny issues. Were our ancestors idiots? Is cognitive ability really so malleable under environmental influence (contrary to what is found in recent twin studies)? Is the principal component identified as g actually time dependent?

\smallskip
\noindent My thoughts are as follows.

(a) Dramatic gains are seen only in certain areas of intelligence, which are plausibly the areas in which modern life provides much more stimulation.

(b) The average person 100 years ago {\it was massively deprived} by today's standards -- much more so than we would ever be allowed to reproduce in a modern twin study. US GDP per capita is 10x higher now, and average years of schooling has increased dramatically. It is not surprising that a child who only received, e.g., 6 years of formal schooling would be far behind someone with 12. In the America of 1900, adults had an average of about 7 years of schooling, a median of 6.5 years, and 25 percent had completed 4 years or less. Modern twin and adoption studies only include individuals raised in a much smaller range of environments -- almost all participants in recent studies have had legally mandated educations, which in the US includes at least several years of high school.

\begin{figure}[tbph]
\begin{center}
\includegraphics[width=10cm]{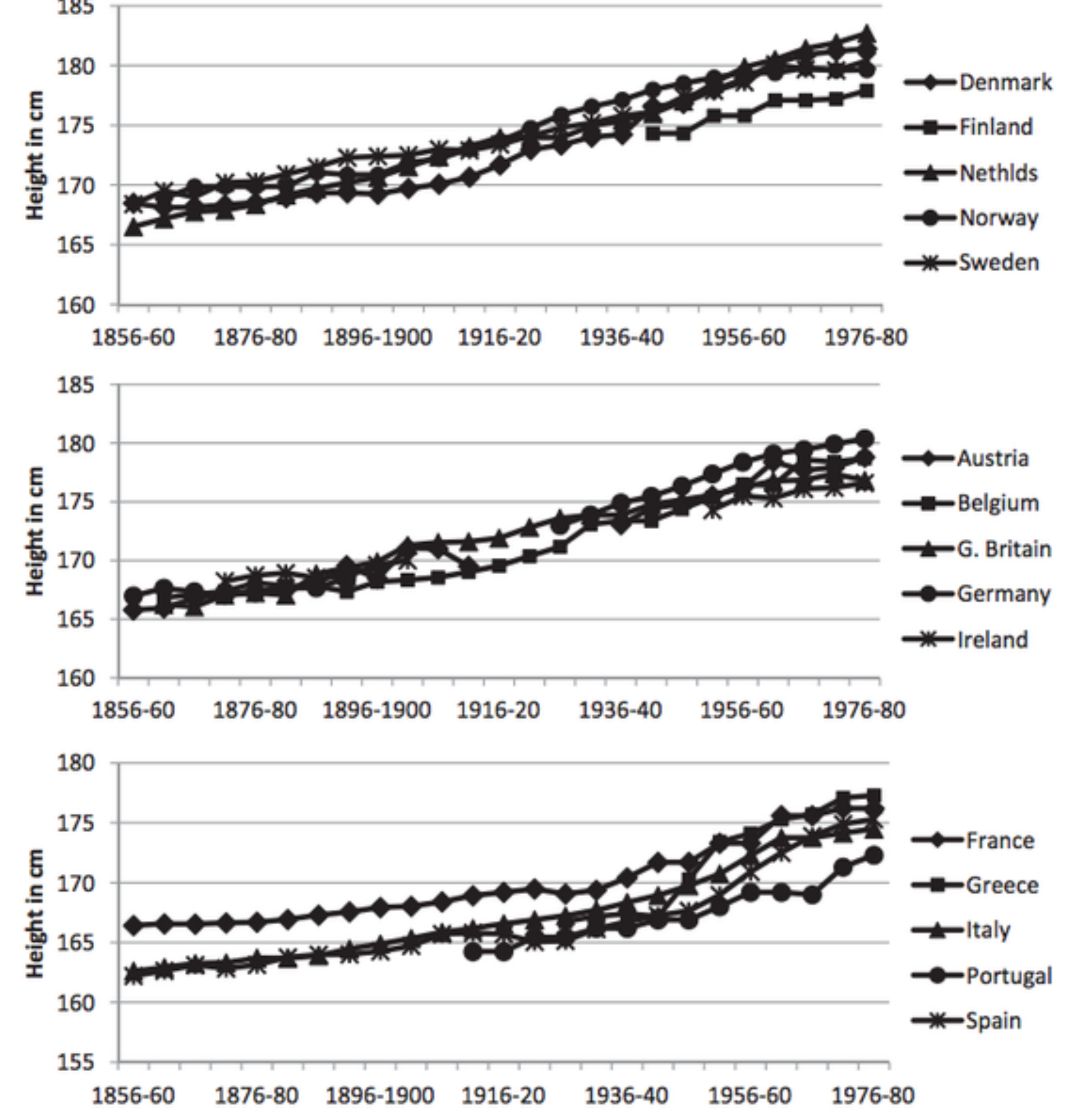}
\end{center}
\caption{Increase in stature in European countries over time, almost +2 SD. Nutrition, hygiene, and average number of years of schooling all improved dramatically over the last 100 years, leading to improvements in both physical and mental development.}
\label{flynn}
\end{figure}

(b) The analogy with height is revealing. While taller parents tend to have taller children (i.e., height is heritable), significant gains in average height {\it which mirror the Flynn effect} (amounting to an almost +2 SD change) have been observed as nutrition and diet have improved. See Fig.(\ref{flynn}).

(d) Variance in adult IQ must have been larger in the past. Figures such as Newton or Thomas Jefferson obviously had tremendously more exposure to ideas and abstract thinking than someone raised on a farm with little or no education, few books, no TV, and no radio. The Flynn effect does not imply that the great geniuses of the past were necessarily inferior to those of today.

\smallskip
\noindent Flynn on the Flynn effect \cite{Flynn1}:
\begin{quote}
... The WISC subtests measure a variety of cognitive skills that are functionally independent and responsive to changes in social priorities over time. The inter-correlations that engender "g" are binding only when comparing individuals within a static social context.

Asking whether IQ gains are intelligence gains is the wrong question because it implies all or nothing cognitive progress. The 20th century has seen some cognitive skills make great gains, while others have been in the doldrums. To assess cognitive trends, we must dissect "intelligence" into solving mathematical problems, interpreting the great works of literature, finding on-the-spot solutions, assimilating the scientific world view, critical acumen, and wisdom.

Our ancestors in 1900 were not mentally retarded. Their intelligence was anchored in everyday reality. We differ from them in that we can use abstractions and logic and the hypothetical to attack the formal problems that arise when science liberates thought from concrete referents. Since 1950, we have become more ingenious in going beyond previously learned rules to solve problems on the spot.

At a given time, genetic differences between individuals (within a cohort) are dominant but only because they have hitched powerful environmental factors to their star. Trends over time (between cohorts) liberate environmental factors from the sway of genes and once unleashed, they can have a powerful cumulative effect.
\end{quote}
Let me reiterate that within a range of favorable environments (i.e., providing good nutrition, hygiene, and access to education), evidence strongly supports the claim that individual differences in cognitive ability are largely associated with genetic differences. However, this does not by itself imply that group differences in cognitive scores are due to genetic causes. Because of our difficult history with race, it would be wise to thoroughly investigate differences between environments experienced by different groups as well as any other confounding factors before arriving at conclusions about genetic causes.

\subsection{Exceptional ability: the far tail}

Many studies show that high cognitive ability is a necessary but not sufficient requirement for scientific achievement. Most research scientists score at the +2 SD level (top few percent) or higher, but there is some evidence that exceptional scientists, especially in theoretical disciplines, have even higher scores.

Harvard psychologist Anne Roe studied 64 randomly selected eminent scientists (ages roughly 40-50) in her 1952 book The Making of a Scientist \cite{Roe}. Among these scientists were physicists Luis Alvarez, Julian Schwinger, Wendell Furry, J.H. Van Vleck and Philip Morse, anthropologist Carleton Coon, psychologist B.F. Skinner, chemist Linus Pauling and geneticist Sewall Wright. Roe administered a high ceiling psychometric test to each scientist, obtaining median scores in both the mathematical and verbal categories in the +4 SD (better than 1 in 10k) range. Thus, randomly sampled eminent scientists were found to be far outliers even among research scientists.

\begin{figure}[tbph]
\begin{center}
\includegraphics[width=12cm]{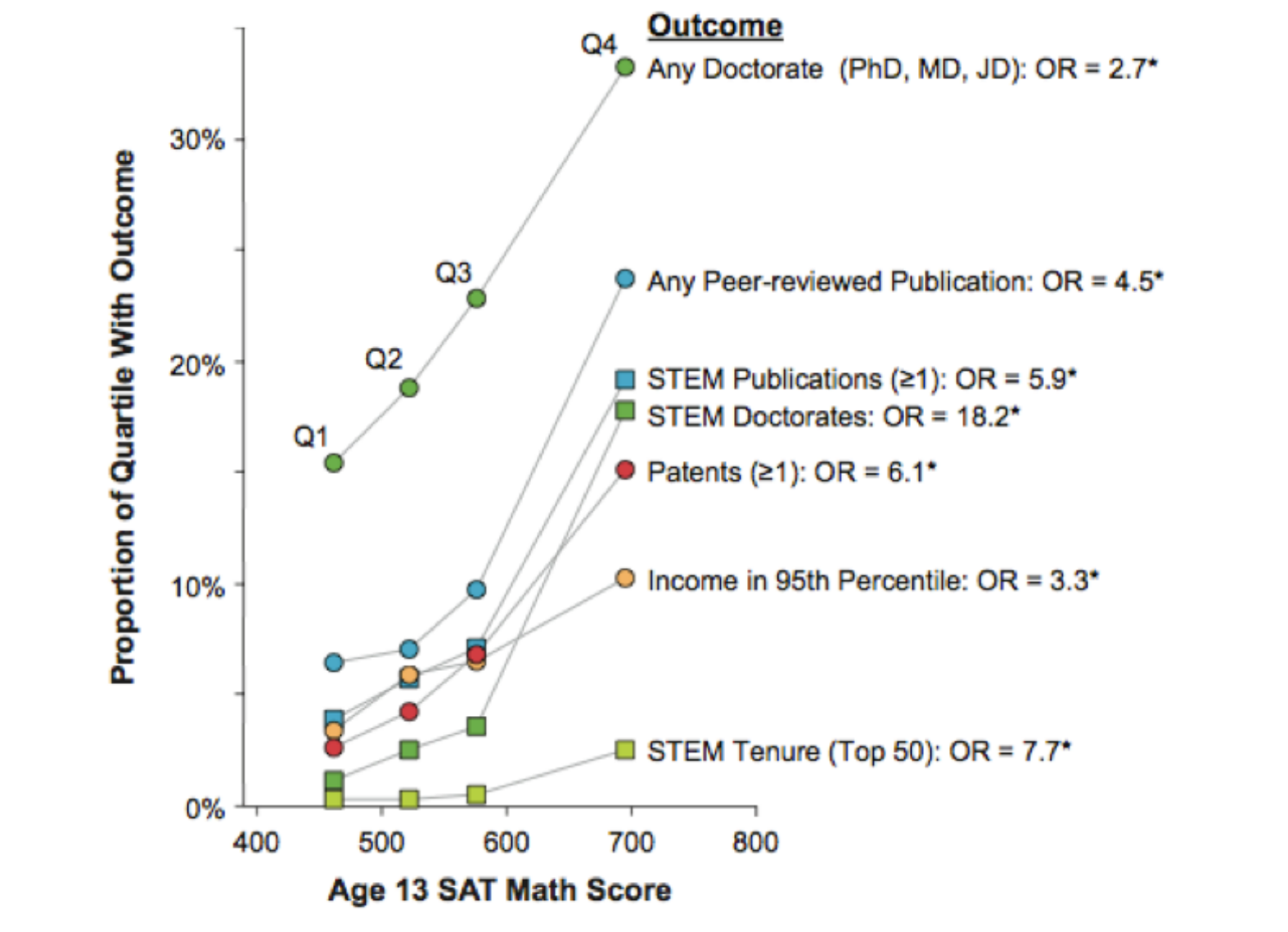}
\end{center}
\caption{Math ability measured at age 13 and life outcomes in a population of gifted (top 1 percent) individuals. The Q4 subgroup scored above the 1 in 10k level of ability.}
\label{smpy}
\end{figure}

A much larger longitudinal study has been conducted of mathematically and verbally precocious individuals \cite{SMPY}. This study followed gifted students from age 13 into middle age. The qualification cutoff was roughly top percentile ability (so each participant in the study is intellectually gifted), while the top subgroup (indicated as Q4 in Fig.(\ref{smpy})) scored at the 1 in 10k level or above. Cognitive ability measured at age 13 was shown to be a strong predictor of future success (probability of earning a terminal degree, publishing a literary or scientific work, earning tenure at a research university, or a patent, etc.), with the top subgroup (Q4) strongly outperforming the lowest one (Q1). Fig.(\ref{smpy}) displays odds ratios as a function of age 13 cognitive ability. I consider the SMPY study to be one of the best arguments for validity of g even in the tail of ability: a short test administered at age 13 or before yields nontrivial life outcome predictions even within the top percentile of talent.

A reasonable surmise from the available evidence is that high cognitive ability is a necessary but certainly not sufficient condition for success in scientific research. Returns to increased ability seem to be positive well into the far tail.

\section{Genetic models}

Below we describe a general model for quantitative phenotypes such as cognitive ability, or height. Let $y$ = individual phenotype, $g_i$ = individual genotype (e.g., list of 1M SNPs or 3B loci), $x_i$ = linear effect sizes, $z_{ij}$ = tensors of nonlinear effect sizes, $\epsilon$ = environmental or other noise.
\begin{equation}
\label{gmodel}
y = \sum_i g_i x_i  + \sum_{ij} g_i g_j z_{ij} + {\cal O} (g^3) + \epsilon
\end{equation}
Terms such as $g_i g_j z_{ij}$ allow one locus to influence another, and also for the effect of a single locus ($i = j$) to behave nonlinearly. We do not include genetic-environmental interaction terms, but rather interpret the equation as describing genetic effects after averaging over some ensemble of (largely favorable) environments.

As we discuss below, there is reason to believe that the linear terms in (\ref{gmodel}) dominate the genetic variance. This simplifies the problem of extracting model parameters $x, z$ from data, although as we show in the section on Compressed Sensing, even the case with significant nonlinear variance can be handled with modern techniques.

\subsection{Approximate additivity: why are phenotype differences approximately linear functions of genotype?}

In studies of twins, siblings and families (see, e.g., Fig.(\ref{twins1})), average similarity in phenotype is approximately linear in degree of relatedness or average fraction of genes shared. This suggests a linear model of gene effects. That is, if each gene has a small additive effect on the phenotype $y$, then the pairwise similarity in $y$ will be directly proportional to the fraction of genes in common between two individuals. Deviations from this direct proportionality are small, suggesting, but not proving, that nonlinear effects (i.e., due to the ${\cal O} (g^2)$ and higher order terms in (\ref{model})) are small.

There are independent arguments, both theoretical and empirical, suggesting that the linear approximation to (\ref{model}), involving only the effect sizes $x_i$, may be fairly accurate. On the empirical side, predictive modeling in animal breeding is based on linear models, and accuracy in commercially important areas such as SNP based prediction of corn or dairy cattle phenotypes \cite{breeding} suggests that similar techniques are likely to work for humans. Below we first review the theoretical arguments for approximate linearity.

%\subsection{Evolutionary dynamics}

Fisher's Fundamental Theorem of natural selection states that the rate of increase of fitness is approximately equal to the {\it additive} (linear) genetic variance \cite{Shraiman,Crow}:
\begin{equation}
\label{Fisher}
\frac{ d \langle F \rangle} {dt} \approx  \sigma_A^2~~.
\end{equation}
This result applies to sexually reproducing species with recombination (generational) timescale smaller than evolutionary timescale. The heuristic justification is as follows. Suppose that a phenotype under selection is controlled both by simple additive effects (i.e., a number of loci whose alleles {\it independently} affect the phenotype either positively or negatively) and also by more complex nonlinear effects (i.e., sets of {\it interacting} loci whose contributions to the phenotype depend on the values of other alleles in the set). Because sexual reproduction scrambles genomes through recombination, adaptations of the complex nonlinear kind are difficult to pass on to offspring. Unless {\it both} mates possess the complex adaptation, the genomes of the descendants may not exhibit the adaptive combination of alleles in the set of multiple loci. On the other hand, additive effects can be reliably passed on because the effect of each individual allele is independent of other loci in the genome. Thus, according to the theorem, response to selection is mainly accomplished through an increase in frequencies of additive alleles that move the phenotype in the fitness increasing direction. Response via nonlinear adaptations requires more time, and hence contributes a subleading term to (\ref{Fisher}).

\begin{figure}[H]
\begin{center}
\includegraphics[width=12cm]{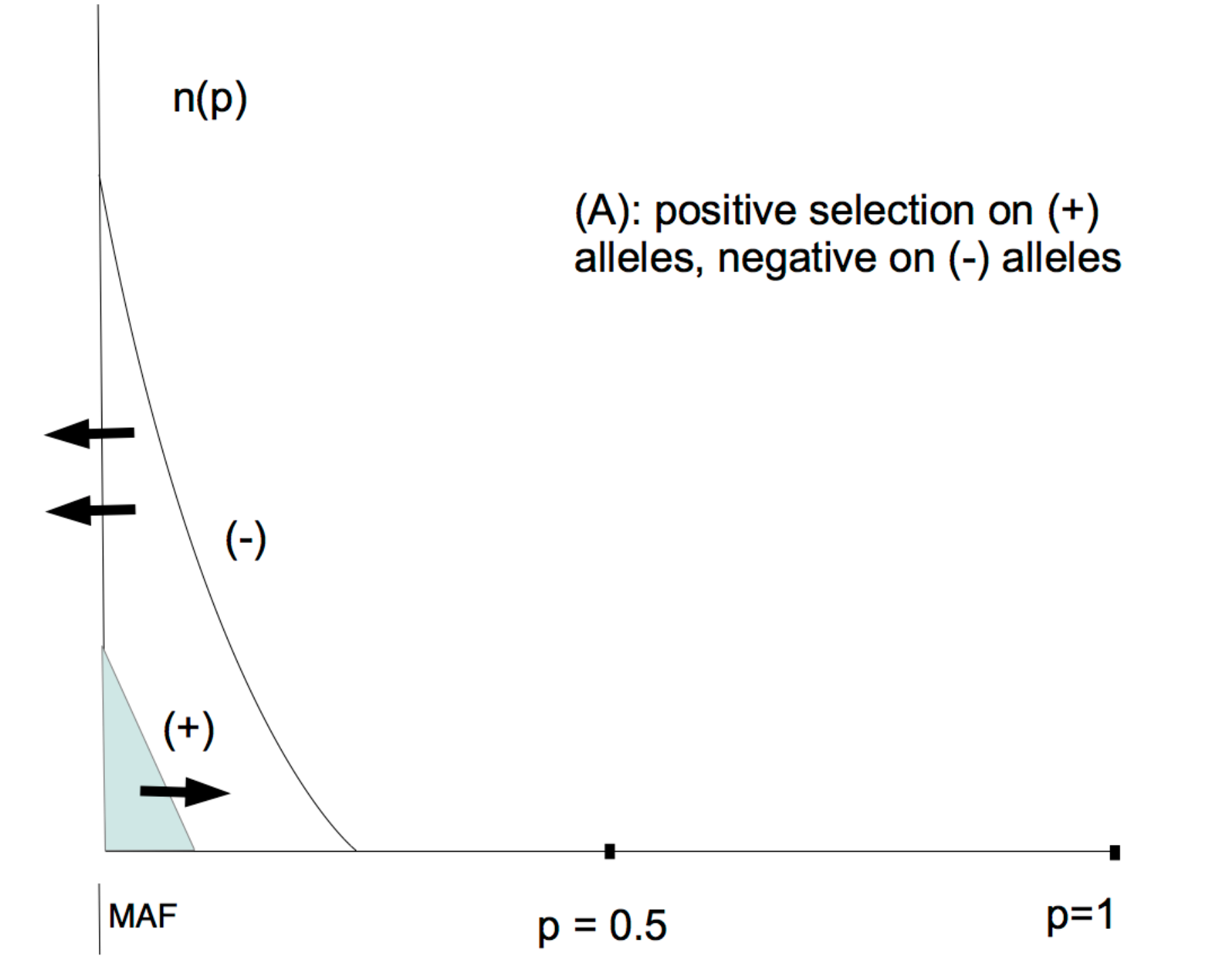}
\includegraphics[width=12cm]{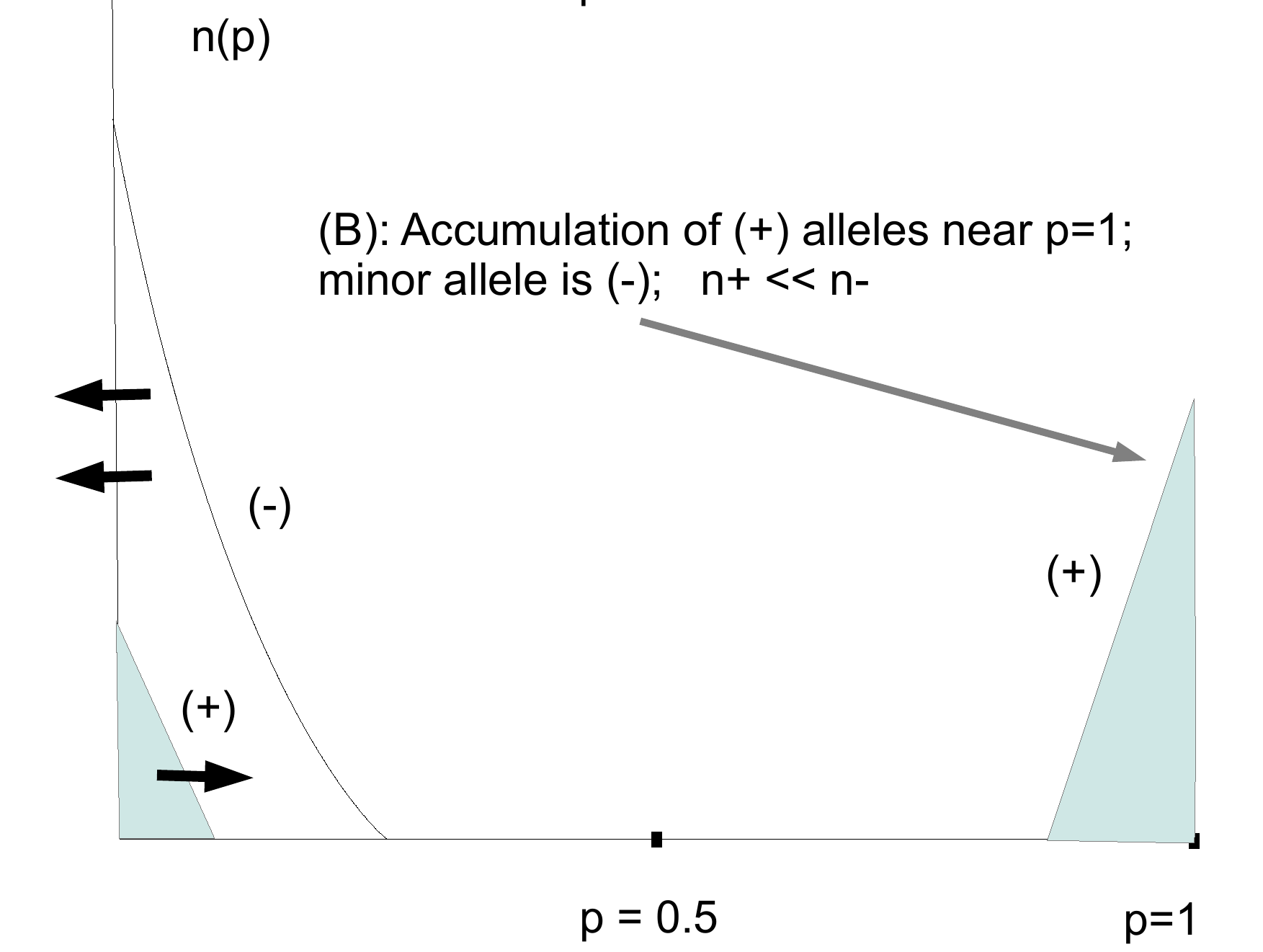}
\caption{The horizontal axis in each figure represents mutant allele frequency (MAF) and the vertical axis represents the density of loci at that frequency. (+) alleles are fitness increasing while (--) alleles are fitness-reducing. Natural selection leads to an increase in frequency for the (+) alleles (i.e., movement to the right) and a decrease in frequency for the (--) alleles (i.e., movement to the left). Over time, (+) allele frequencies are driven to larger MAF (even, to fixation or MAF = 1), while (--) allele frequencies are driven to low values (perhaps even zero). After a long period of selection, we expect to find fitness-reducing (deleterious) alleles at low MAF in the population.}
\end{center}
\label{evofig}
\end{figure}

Fig.(\ref{evofig}) helps to visualize how allele frequencies shift in response to selection. The horizontal axis in each figure represents mutant allele frequency (MAF) and the vertical axis represents the density of loci at that frequency. (+) alleles are fitness increasing while (--) alleles are fitness reducing. 

\begin{figure}[H]
\begin{center}
\includegraphics[width=12cm]{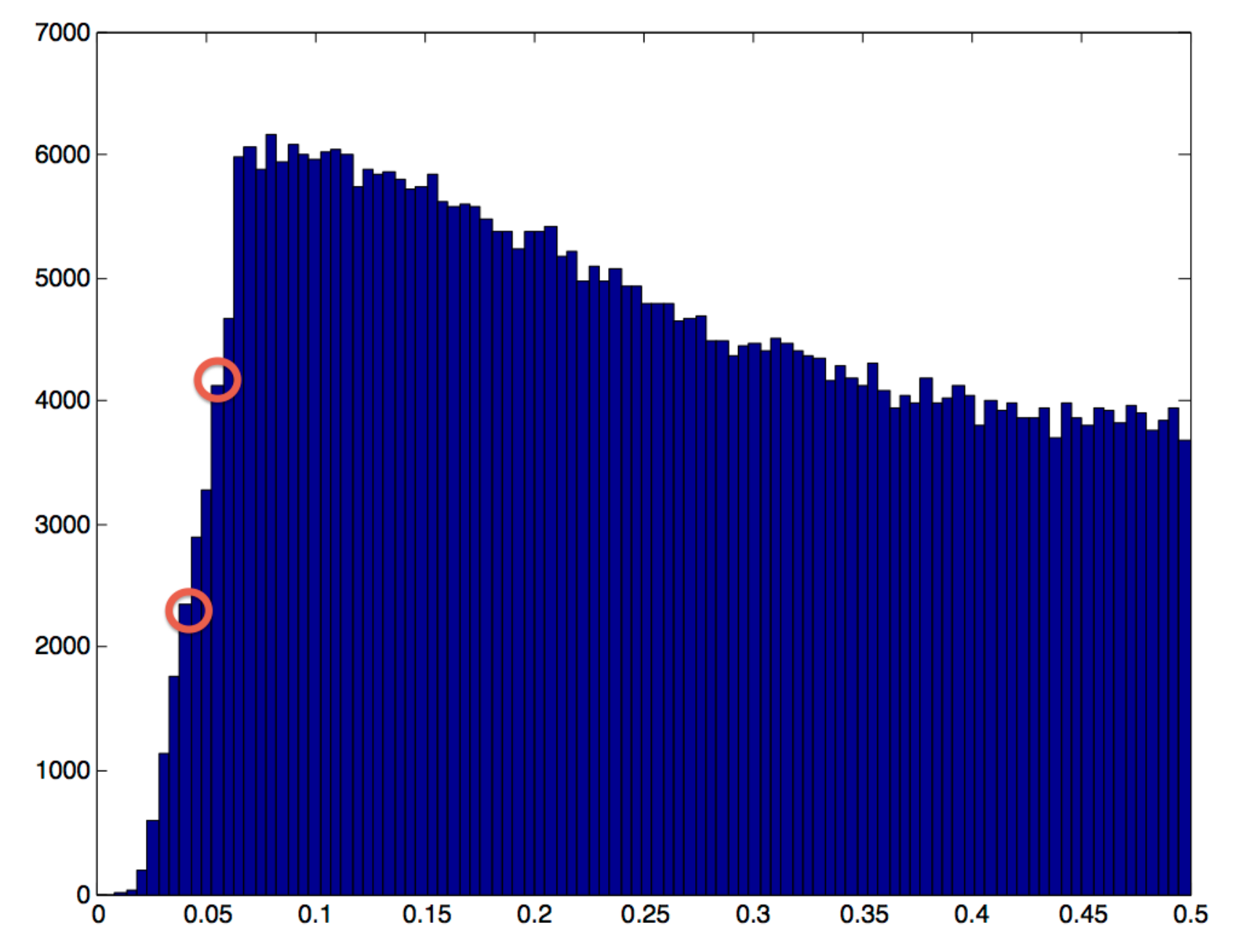}
\includegraphics[width=12cm]{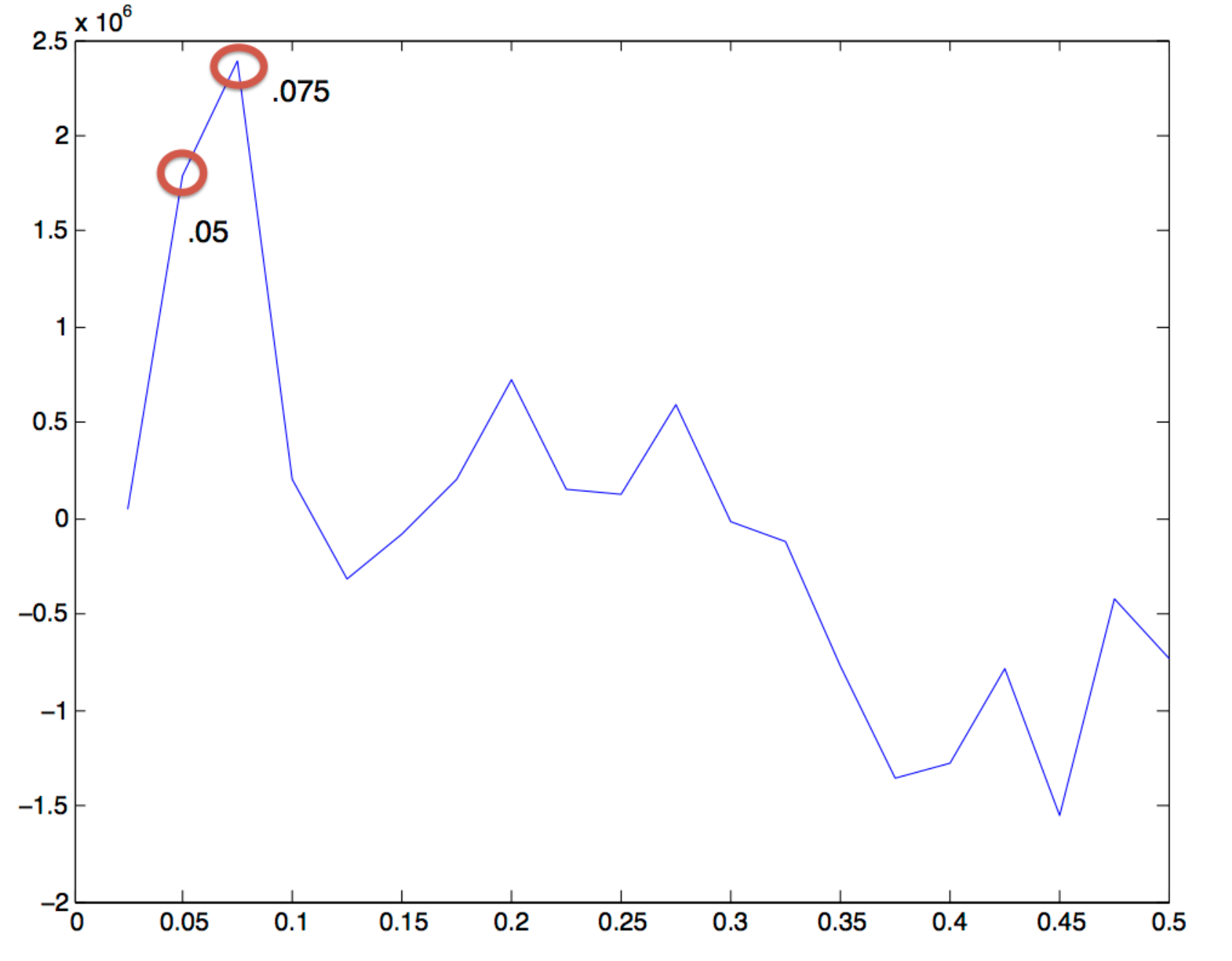}
\end{center}
\caption{SNPs contributing to phenotype-correlated genetic distance are concentrated at small minor allele frequency. Top panel shows density of SNPs on chip by minor allele frequency. Lower panel shows contribution to genetic distance binned by minor allele frequency.}
\label{histo}
\end{figure}

Natural selection leads to an increase in frequency for the (+) alleles (i.e., movement to the right) and a decrease in frequency for the (--) alleles (i.e., movement to the left). Over time, (+) allele frequencies are driven to larger MAF (even, to fixation or MAF = 1), while (--) allele frequencies are driven to low values (perhaps even zero). After a long period of selection, we expect to find fitness-reducing (deleterious) alleles at low frequency in the population.

Given the significant increase in hominid intelligence and stature over millions of years, it seems reasonable to assume that both have been under positive selection for extended periods of time. This suggests that frequencies for causal variants related to height and cognitive ability will typically have extreme values. Increased density at the extremes is also predicted by a model where the trait is selectively neutral, but natural selection will tend to exaggerate this tendency.

A consequence of this observation is that nonlinear effects in phenotype differences will be suppressed \cite{Visscher}. To see this, consider diploid genotypes:  CC, cC, cc, where C is the common allele and c the rare allele. A simple example of a nonlinear interaction (also referred to as {\it epistasis}): the effect of cc may not be twice effect of cC. Alternatively, we might have a multi-locus interaction such that the effect of a c variant depends on the state of another locus, DD, dD, dd. In either scenario, if the probabilities of a c or d variant are both small, we can expand the phenotype function to lowest order in the rare variants c and d, and neglect cases where cc or dd are realized. If the minor allele frequency $\sim p$ for both c and d, the prevalence of cc and dd is only $\sim p^2$. Since many loci affect a polygenic trait, {\it most} of the difference between two individuals on the trait will be due to differences such as DD vs Dd. Up to subleading corrections of order $p$, the sum of gene effects can be {\it approximated} by a linear function of genotype. Note, this observation does not assume a linear genetic architecture -- nonlinear (epistatic) effects at the single or multiple gene level are permitted. However, the {\it aggregate} effect of many genes will be approximately linear as long as minor allele frequencies are small. A high degree of non-linearity at the genetic level can still correspond to almost linear aggregate variation between two individuals.

The preceding discussion is not intended to convey an overly simplistic view of genetics or systems biology. Complex nonlinear genetic systems certainly exist and are realized in every organism. However, quantitative differences {\it between} individuals within a species may be largely due to independent linear effects of specific genetic variants. As noted, linear effects are the most readily evolvable in response to selection, whereas nonlinear gadgets are more likely to be fragile to small changes. (Evolutionary adaptations requiring significant changes to nonlinear gadgets are improbable and therefore require exponentially more time than simple adjustment of frequencies of alleles of linear effect.) One might say that to first approximation, {\it Biology = linear combinations of nonlinear gadgets}, and most of the variation between individuals is in the (linear) way gadgets are combined, rather than in the realization of {\it different gadgets} in different individuals.

Linear models works well in practice, allowing, for example, SNP-based prediction of quantitative traits (milk yield, fat and protein content, productive life, etc.) in dairy cattle. From P. VanRaden et al., Invited Review: Reliability of genomic predictions for North American Holstein bulls \cite{breeding}:
\begin{quote}
Marker effects for most ... traits were evenly distributed across all chromosomes with only a few regions having larger effects, which may explain why the infinitesimal model and standard quantitative genetic theories have worked well. The distribution of marker effects indicates primarily polygenic rather than simple inheritance and suggests that the favorable alleles will not become homozygous quickly, and genetic variation will remain even after intense selection. Thus, dairy cattle breeders may expect genetic progress to continue for many generations.
\end{quote}
Agricultural animals are highly inbred, leading to far fewer independent haploblocks in their DNA (i.e., lower genetic diversity). This makes predictive modeling much easier than in the case of humans. Nevertheless, the observation that additive effects predominate should generalize to other species.

\subsection{Estimate of number of causal variants}

Using a simple additive model one can roughly estimate the number of causal variants. Assume that there are $N$ causal variants with typical minor allele frequency $p \ll1$ (the minor allele is defined as having lower frequency, regardless of whether it is mutant or ancestral). Let the minor alleles (--) each have equal small negative effect on the phenotype. From the binomial theorem we know that the distribution of number of minor alleles in the population is approximately Gaussian with standard deviation SD $\approx ( p N )^{1/2}$. That is, to cause a one SD shift in the phenotype requires this number of allele flips from (+) to (--) or vice-versa. Averaging over pairs of randomly selected individuals, the pairwise SNP distance (number of SNP differences between them) should increase as the phenotype difference increases. The rate of this increase (number of SNPs per population SD of change in the phenotype) yields an estimate for $( p N )^{1/2}$

We performed this calculation on samples of several thousand individuals for whom we had SNP genotypes and phenotype values for height and cognitive ability. Our estimate for the number of SNPs per SD of phenotype difference in both cases was roughly 40. Taking $p \approx 0.1 - 0.2$ (see Fig.(\ref{histo})), this implies $N \sim 10$k, or roughly ten thousand causal variants. Unfortunately, the statistical noise was quite large for this method: pairwise distance for this population is 261k $\pm$ 1.5k SNPs. So the method attempts to detect SNP distances $\sim 40$ on a background of fluctuations $\sim \pm 1500$. Given $10^6$ {\it independent} pairs of individuals this would be possible through averaging (thereby suppressing the statistical error by $\sim 1 / \sqrt{10^6}$). However with a few thousand unique individuals the effective number of independent pairs is much smaller. Using a bootstrap or jackknife method (analyzing random subsets of the data, and estimating how the noise fluctuations scale with sample size), we estimated that background fluctuations could be suppressed to the level of $\sim (50-100)$ SNPs, which is somewhat larger than the signal we detected. Nevertheless, our results suggest an {\it upper bound} on $N$ which is not much larger than 10k. That is, if $N$ were much larger than this, we would have seen a clean signal above background noise. A lower bound on $N$ results from GWAS: the limited success of smaller studies puts an upper bound on typical effect sizes of causal alleles; small effect sizes imply a large number of causal variants.

Interestingly, in our data (Fig.(\ref{histo})) the alleles contributing to the SNP distance associated with phenotype variation are concentrated mostly at small minor allele frequency.

\begin{figure}[tbph]
\begin{center}
\includegraphics[width=11cm]{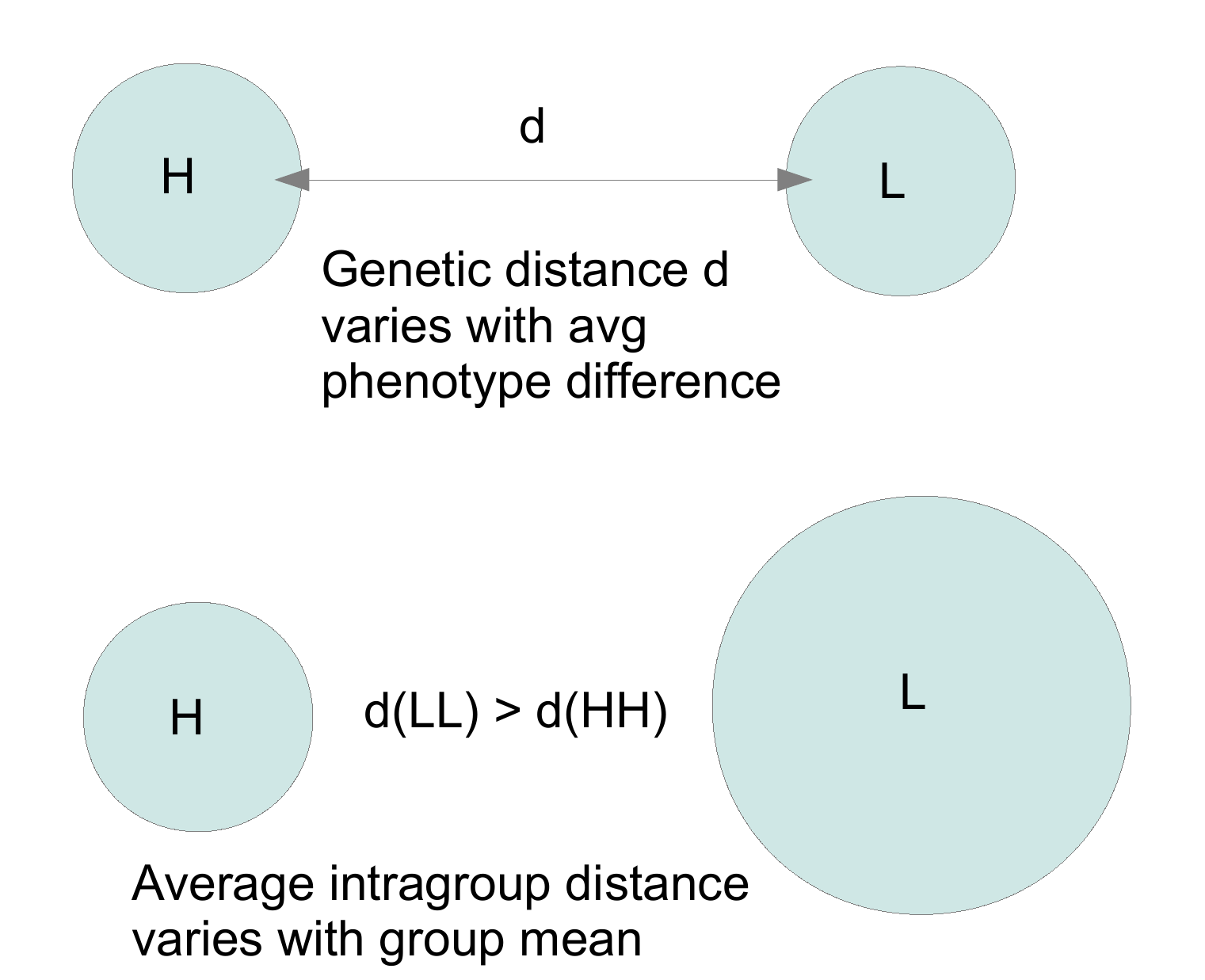}
\end{center}
\caption{Pairwise genetic distance is correlated with phenotype difference and with average phenotype. This allows a rough estimate of the number of causal variants.}
\label{distances}
\end{figure}

We also noted that pairwise distances changed systematically with {\it average} phenotype value in the pair (Fig.(\ref{distances})). Pairs of individuals who were both below average in stature or cognitive ability tended to have more SNP changes between them than pairs who were both above average. This result supports the assumption that the minor allele (--) tends to reduce the phenotype value. In a toy model with, e.g., $p = 0.1, N= 10$k, an individual with average phenotype would have 9k (+) variants and 1k (--) variants. A below average ($\approx - 3$ SD) person might instead have 1100 (--) variants, and an above average individual ($\approx + 3$ SD) 900 (--) variants. The typical SNP distance between genotypes with 1100 (--) variants is larger than that for genotypes with 900 (--) variants, as there are many places to place the (--) alleles in a list of 10k total causal variants. Two randomly chosen individuals will generally not overlap much in the positions of their (--) variants, so each additional (--) variant tends to increase the distance between them.

\newpage

\subsection{An evolution beyond man?}

\begin{quote}
{\it I have known a great many intelligent people in my life. I knew Planck, von Laue and Heisenberg. Paul Dirac was my brother in law; Leo Szilard and Edward Teller have been among my closest friends; and Albert Einstein was a good friend, too. But none of them had a mind as quick and acute as Jansci [John] von Neumann. I have often remarked this in the presence of those men and no one ever disputed me.} -- Nobel Laureate Eugene Wigner
\smallskip

{\it One of his remarkable abilities was his power of absolute recall. As far as I could tell, von Neumann was able on once reading a book or article to quote it back verbatim; moreover, he could do it years later without hesitation. ... On one occasion I tested his ability by asking him to tell me how The Tale of Two Cities started. Whereupon, without any pause, he immediately began to recite the first chapter and continued until asked to stop after about ten or fifteen minutes.} -- Herman Goldstine, mathematician and computer pioneer.
\end{quote}

There is good evidence that existing genetic variants in the human population (i.e., alleles affecting intelligence that are found today in the collective world population, but not necessarily in a single person) can be combined to produce a phenotype which is far beyond anything yet seen in human history. This would not surprise an animal or plant breeder -- experiments on corn, cows, chickens, drosophila, etc. have shifted population means by many standard deviations relative to the original wild type.

Take the case of John von Neumann, widely regarded as one of the greatest intellects in the 20th century, and a famous polymath. He made fundamental contributions in mathematics, physics, nuclear weapons research, computer architecture, game theory and automata theory. In addition to his abstract reasoning ability, von Neumann had formidable powers of mental calculation and a photographic memory. In my opinion, genotypes exist that correspond to phenotypes as far beyond von Neumann as he was beyond a normal human.

The quantitative argument for why there are many SD's to be had from tuning genotypes is straightforward. Suppose variation in cognitive ability is

1. highly polygenic (i.e., controlled by $N$ loci, where $N$ is large, such as 10k), and

2. approximately linear (note the additive heritability of g is larger than the non-additive part).

Then the population SD for the trait corresponds to an excess of roughly $N^{1/2}$ positive alleles (for simplicity we suppress dependence on minor allele frequency). A genius like von Neumann might be +6 SD, so would have roughly $6 N^{1/2}$ more positive alleles than the average person (e.g., $\sim 600$ extra positive alleles if $N$ = 10k). But there are roughly $+N^{1/2}$ SDs in phenotype ($\sim 100$ SDs in the case $N \sim$ 10k) to be had by an individual who has essentially {\it all} of the $N$ positive alleles! As long as $N^{1/2} \gg 6$, there is ample extant variation for selection to act on to produce a type superior to any that has existed before. The probability of producing a "maximal type" through random breeding is exponentially small in $N$, and the historical human population is insufficient to have made this likely.

The content of this basic calculation underlies the work of animal and plant breeders. As leading population geneticist James Crow of Wisconsin wrote \cite{Crow}:
\begin{quote}
The most extensive selection experiment, at least the one that has continued for the longest time, is the selection for oil and protein content in maize (Dudley 2007). These experiments began near the end of the nineteenth century and still continue; there are now more than 100 generations of selection. Remarkably, selection for high oil content and similarly, but less strikingly, selection for high protein, continue to make progress. There seems to be no diminishing of selectable variance in the population. The effect of selection is enormous: the difference in oil content between the high and low selected strains is some 32 times the original standard deviation.
\end{quote}
To take another example, wild chickens lay eggs at the rate of roughly one per month. Domesticated chickens have been bred to lay almost one egg per day. (Those are the eggs we have for breakfast!) Of all the wild chickens in evolutionary history, probably not a single one produced eggs at the rate of a modern farm chicken.

\section{Compressed Sensing and the genome}

My recent interest in genomics is largely due to rapid advances in genotyping technologies and the consequent possibility of dramatic progress. As a student I had judged quantitative genetics to be interesting but too distant from rigorous confrontation with experiment to warrant significant effort (at least on my part). But the situation has changed radically, and deep problems in evolutionary theory and biology can now be attacked with some confidence of results in the foreseeable future.

The questions addressed in this section were among the first that occurred to me upon reengaging with genetic science, and are, at least in my view, some of the most important: Are problems of genomic prediction tractable in the foreseeable future? What are the most efficient computational methods and their performance characteristics? How much predictive power can we expect to obtain for quantitative traits at a given heritability, as a function of the amount of available data and computational power? 

In this section I describe new methods for extraction of genomic model parameters from data. As a consequence of these advances, we can make rough estimates for the amount of computation and data required to build accurate phenotype predictors of the type already used in animal breeding \cite{breeding}.

\subsection{Application to linear genetic models}

Compressed Sensing (CS) allows efficient solution of underdetermined linear systems:  
\begin{equation}
\label{ax}
y = Ax + \epsilon ,  
\end{equation}
($\epsilon$ is a noise term) using a form of penalized regression. L1 penalization, or LASSO, involves minimization of an objective function over candidate vectors $\hat{x}$:
\begin{equation}
\label{O}
O = \vert \vert y - A \hat{x} \vert \vert_{L2} + \lambda \vert \vert \hat{x} \vert \vert_{L1}~~,
\end{equation}
where the penalization parameter is determined by the noise variance. Because $O$ is a convex function it is easy to minimize. Recent theorems \cite{CS} provide performance guarantees, and show that the $\hat{x}$ that minimizes $O$ is overwhelmingly likely to be the sparsest solution to (\ref{ax}). In the context of genomics, $y$ is the phenotype, $A$ is a matrix of genotypes (in subsequent notation we will refer to it as $g$), $x$ a vector of effect sizes, and the noise is due to nonlinear gene-gene interactions and the effect of the environment. 

Let $p$ be the number of variables (i.e., dimensionality of $x$, or number of genetic loci), $s$ the sparsity (number of variables or loci with nonzero effect on the phenotype; i.e., nonzero entries in $x$) and $n$ the number of measurements of the phenotype (i.e., dimensionality of $y$ or the number of individuals in the sample). Then  $A$  is an  $n \times p$  dimensional matrix. Traditional statistical thinking suggests that $n > p$  is required to fully reconstruct the solution  $x$  (i.e., reconstruct the effect sizes of each of the loci). But recent theorems in compressed sensing show that  $n > C s \log p$ (for constant $C$ defined over a class of matrices $A$) is sufficient if the matrix $A$ has the right properties (is a good compressed sensor). These theorems guarantee that the performance of a compressed sensor is nearly optimal -- within an overall constant of what is possible if an oracle were to reveal in advance which $s$  loci out of $p$ have nonzero effect. In fact, one expects a phase transition in the behavior of the method as $n$ crosses a critical threshold $n_{\star}$ given by the inequality. In the good phase ($n > n_{\star}$), full recovery of $x$ is possible.

In \cite{GCS}, it is shown that

a. Matrices of human SNP genotypes are good compressed sensors and are in the universality class of random matrices. The phase behavior is controlled by scaling variables such as $\rho = s / n$, and simulation results predict the sample size threshold for future genomic analyses.

b. In applications with real data the phase transition can be detected from the behavior of the algorithm as the amount of data $n$ is varied. (For example, in the low noise case the mean P-value of selected, or non-zero, components of $x$ exhibits a sharp jump at $n_{\star}$.) A priori knowledge of $s$ is not required; in fact one deduces the value of $s$ this way.

c.  For heritability $h^2 = 0.5$ and $p \sim 10^6$ SNPs, the value of $C  \log  p \sim 30$. For example, a trait which is controlled by $s = 10$k loci would require a sample size of $n \sim 300$k individuals to determine the (linear) genetic architecture (i.e., to determine the full support, or subspace of nonzero effects, of $x$).

\subsection{Application to nonlinear genetic models}

Realistic models relating phenotype to genotype exhibit nonlinearity (epistasis), allowing distinct regions of DNA to interact with one another. For example, one allele can influence the effect of another, altering its magnitude or sign, even silencing the second allele entirely.  As discussed previously, we should not be surprised to find that the largest component of genetic variance is linear (additive) \cite{Visscher}, but even in this case nonlinear interactions accounting for some smaller component of variance are expected to be present. To obtain the best possible model for prediction of phenotype from genotype, or to obtain the best possible understanding of the genetic architecture, requires the ability to extract information concerning nonlinearity from phenotype--genotype (e.g., GWAS) data. 

It is a common belief in genomics that nonlinear interactions (epistasis) in complex traits make the task of reconstructing genetic models extremely difficult, if not impossible. In fact, it is often suggested that overcoming nonlinearity will require much larger data sets and significantly more computing power. We have developed a nonlinear generalization of Compressed Sensing (CS) and applied it to this class of problems, using both real and simulated SNP genotypes \cite{NLCS}. Our results show that in broad classes of plausibly realistic models, most of the nonlinear as well as linear genetic variance can be recaptured using this technique.

Consider the model (we include explicit indices for clarity; $1 \leq a \leq n$ labels individuals and $1 \leq i,j \leq p$ label genomic loci)
\begin{equation}
\label{model}
y^a = \sum_i g^a_i z_i ~+~  \sum_{ij} g^a_i Z_{ij} g^a_j ~+~ \epsilon^a ~~~,
\end{equation}
where $g$ is an $n \times p$ dimensional matrix of genomes, $z$ is a vector of linear effects, $Z$ is a matrix of nonlinear interactions, and $\epsilon$ is a random error term. We could include higher order (i.e., gene-gene-gene) interactions if desired.

Suppose that we apply conventional CS to data generated from the model above. This is equivalent to finding the best-fit linear approximation
\begin{equation}
\label{yax}
y^a \approx \sum_i g^a_i {x}_i~~.
\end{equation}
If enough data ($n \sim s \log p$, where $s$ is the sparsity of $x$) is available, the procedure will produce the best-fit hyperplane approximating the original data.

It seems plausible that the support of $x$, i.e., the subspace defined by non-zero components of $x$, will coincide with the subset of loci which have nonzero effect in {\it either} $z$ {\it or} $Z$ of the original model. That is, if the phenotype is affected by a change in a particular locus in the original model (either through a linear effect $z$ or through a nonlinear interaction in $Z$), then CS will assign a nonzero effect to that locus in the best-fit linear model (i.e., in $x$). We have verified that this hypothesis is largely correct: the support of $x$ coincides with the support of $(z,Z)$ except in some special cases where nonlinearity masks the role of a particular locus. See \cite{FAE} for theoretical discussion of the causal meaning of the average effect of gene substitution (i.e., linear effect). 

Is it possible to do better than the best-fit linear effects vector $x$? How hard is it to reconstruct both $z$ and $Z$ of the original nonlinear model? This is an interesting problem both for genomics (in which, even if the additive variance dominates, there is likely to be residual non-additive variance) and other nonlinear physical systems.

It is worth noting that although (\ref{model}) is a nonlinear function of $g$ -- i.e., it allows for epistasis, gene-gene interactions, etc. -- the phenotype $y$ is nevertheless a {\it linear} function of the parameters $z$ and $Z$. One could in fact re-express (\ref{model}) as
$$
y^a = \sum_i G^a_i (g) X_i + \epsilon^a
$$
where $X$ is a vector of effects (to be extracted) and $G$ the most general nonlinear function of $g$ over the $s$-dimensional subspace selected by the first application of CS resulting in (\ref{yax}). Working at, e.g., order $g^2$, $X$ would have dimensionality $s(s-1)/2 + 2s$, enough to describe all possible linear and quadratic terms in (\ref{model}). 

Given the random nature of $g$, it is not surprising that $G$ is also a well-conditioned CS matrix (we have verified that this is the case). Potentially, the number of nonzero components of $X$ could be $\sim s^k$ at order $g^k$. However, if the matrix $Z$ has a sparse or block-diagonal structure (i.e., individual loci only interact with some limited number of other genes, not all $s$ loci of nonzero effect; this seems more likely than the most general possible $Z$), then the sparsity of $X$ is of order a constant $k$ times $s$. Thus, extracting the full nonlinear model is only somewhat more difficult than the $Z = 0$ case. Indeed, the data threshold necessary to extract $X$ scales as $\sim k s \log (s(s-1)/2 + 2s)$, which is less than $s \log p$ as long as $k \log (s(s-1)/2 + 2s) < \log p$.
\bigskip

The process for extracting $X$, which is equivalent to fitting the full nonlinear model in (\ref{model}), is as follows:
 
1. Run CS on $(y,g)$ data, using linear model (\ref{yax}). Determine support of $x$: subset defined by $s$ loci of nonzero effect.

2. Compute $G(g)$ over this subspace. Run CS on $y = G(g) \cdot X$ model to extract non-zero components of $X$. These can be translated back into the linear and nonlinear effects of the original model (i.e., nonzero components of $z$ and $Z$).

We find that in many cases steps 1 and 2 lead to almost perfect reconstruction of the original model (\ref{model}) given enough data $n$. We have also investigated the following issues.

a. When can nonlinear effects hide causal loci from linear regression (step 1)? In cases of this sort the locus in question would not be discovered by GWAS using linear methods.

b. Both matrices g and G(g) seem to be well-conditioned CS matrices.

c. For a given partition of variance between linear (L), nonlinear (NL) and IID error $\epsilon$, how much data $n_{\star}$ is required before complete selection of causal variants occurs (i.e., crossing of the phase boundary for algorithm performance)? Typically if step 1 is successful then with the same amount of data step 2 will also succeed.

\bigskip

We have tested the proposed method on a variety of plausible nonlinear genetic models, and find that it can recover a significant fraction of the predictive power (equivalently, variance) associated with nonlinear effects. 
%The upper left panels of Fig.(\ref{fig:BD}) and Fig.(\ref{fig:PS}) show that 
In the thousands of models we studied, the method typically recovers half or more of the nonlinear genetic variance. To take a specific example, for nonlinear variance $\sigma^2_\textrm{NL} \sim 0.25$, $\textrm{var}( \epsilon ) = 0.3$, over a third of the total genetic variance $$h^2_\textrm{broad sense} \equiv 1 - \textrm{var}( \epsilon ) = 0.7$$ is due to nonlinear effects. Step 2 of our method recovers all but $\sigma^2_\textrm{R} \sim 0.1$ of the total genetic variance, using the same amount of data as in the linear first step: of order $100 s$. 

The fraction of variance not recovered by our method is largely due to the causal variants that are not detected by step 1 of the algorithm -- i.e., the fraction of zeros. These variants would also escape detection by linear regression using the same amount of sample data.

\section{GWAS results}

Genome Wide Association Studies (GWAS) seek to detect statistical associations between genetic variants and phenotype. These studies require large phenotype$\vert$genotype data sets. A P-value of $5 \times10^{-8}$ is required for an association to be considered genome-wide significant. This criterion results from the usual $p < 0.05$ significance corrected for multiple testing of $\sim 10^6$ independent SNPs in the human genome. Associations at genome wide significance are typically found to replicate reliably in different populations, even of varying ancestry.

\begin{figure}[tbph]
\begin{center}
\includegraphics[width=12cm]{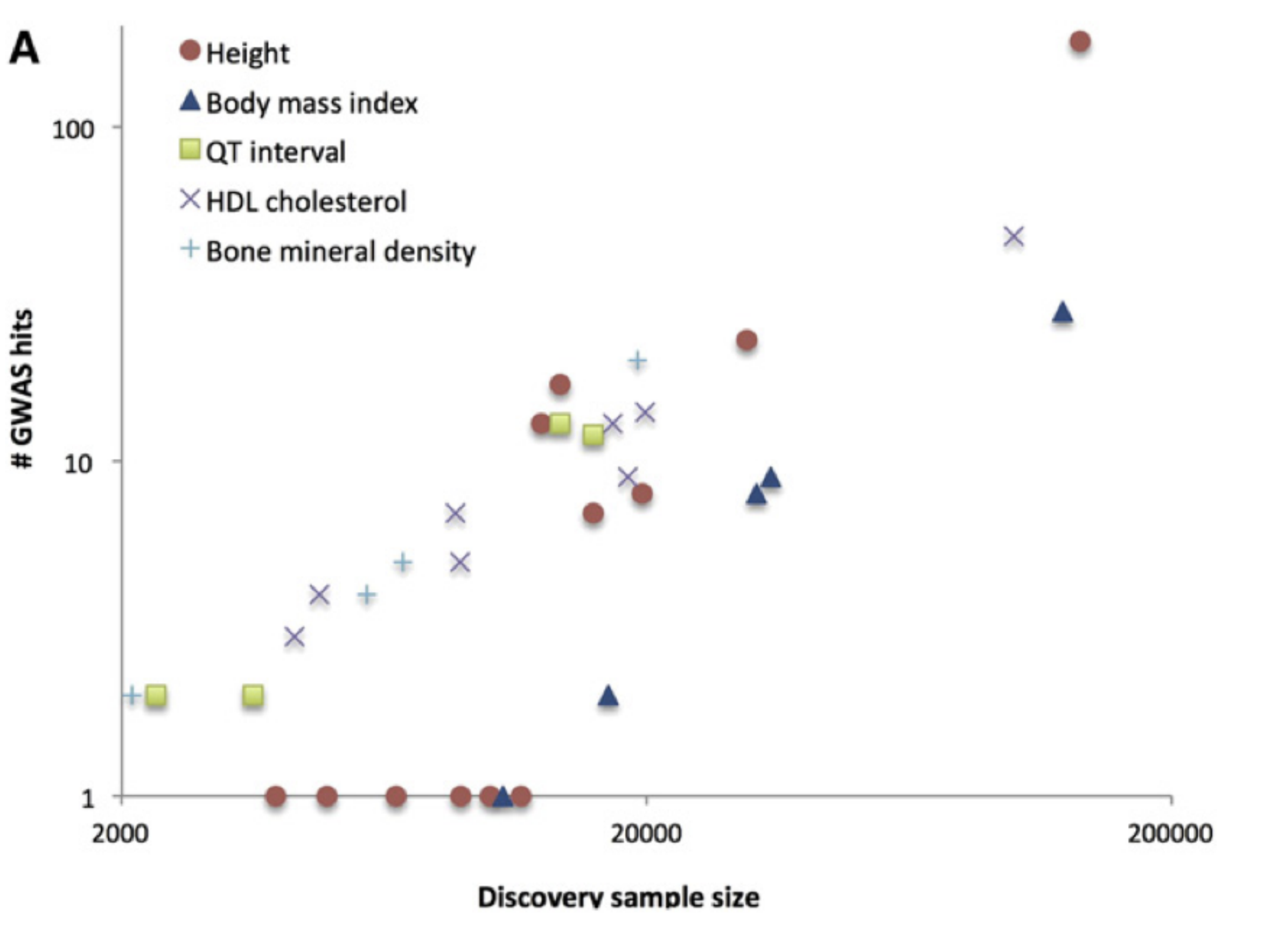}
\end{center}
\caption{GWAS history. Thresholds are observed where sample size reaches the necessary statistical power to discover alleles accounting for largest variance in the population (i.e., the easiest to detect) at genome wide significance ($p < 5 \times10^{-8}$). Once this threshold is passed, steady discovery of new variants tracks increasing sample size.}
\label{gwashistory}
\end{figure}

GWAS discovery is driven by statistical power. Fig.(\ref{gwashistory}) displays number of genome wide significant hits versus sample size for a variety of phenotypes \cite{GWAS}. It seems reasonable to assume that cognitive ability will follow a similar trajectory: once the minimum threshold required to discover the alleles accounting for the largest portion of variance (i.e., the easiest to detect) is exceeded, more and more alleles are discovered with increasing sample size.

Conventional GWAS methodology relies on simple regression of phenotype against a specific variant. Our Compressed Sensing results suggest that the set of associated variants for a trait (i.e., the support of the vector $\hat{x}$ in equation (\ref{O})) can be discovered all at once after a critical threshold of sample size is passed. This sample size is probably of order a million individuals for both height and cognitive ability. Simple extrapolation of the height points in Fig.(\ref{gwashistory}) also suggests that linear regression with millions of genomes will produce thousands of genome wide significant hits.

\subsection{SSGAC}

The SSGAC (Social Science Genetic Association Consortium) is a large collaboration with participant researchers from many dozens of universities in the US and abroad. Their corpus includes over 70 different sample cohorts (accounting for over 100k individuals) for whom data on educational attainment and / or cognitive ability are available. In 2013 they announced genome wide significant hits on 3 SNPs associated with educational attainment \cite{SSGAC}. Subsequent work \cite{SSGAC1} supports the hypothesis (see analysis in Supplement, especially figure S22 and section 7, of \cite{SSGAC}) that these SNPs affect cognitive ability.

\subsection{BGI Project}

The BGI Cognitive Genomics Lab has obtained DNA samples from over 2000 intellectually gifted individuals \cite{BGI}. Whole genome sequences at 4x coverage have been obtained for a subset of these, with the remainder delayed until the new Complete Genomics platform (CG was acquired by BGI in 2013) is operative. Of the gifted individuals in our sample, about half are volunteers who either hold advanced credentials from leading PhD programs in quantitative subjects, or submitted exceptionally high scores on standardized tests such as SAT, ACT or GRE, or both. The remainder of the cohort are alumni of gifted programs similar to SMPY who tested at the 1 in 10k level before age 13 (DNA samples obtained by leading behavior geneticist Robert Plomin of King's College London using funds from the Templeton Foundation).

\subsection{Project Einstein}

Sequencing pioneer Jonathan Rothberg (founder of 454 Life Sciences and Ion Torrent) is funding Project Einstein \cite{Einstein}, which seeks to identify the genetic basis for mathematical genius. DNA samples have been obtained from 400 leading mathematicians and theoretical physicists, with whole genome sequences expected to be obtained by end of summer 2014.

\newpage

\section{The future}
\begin{quote}
{\it Alpha children wear grey. They work much harder than we do, because they're so frightfully clever. I'm really awfully glad I'm a Beta, because I don't work so hard. And then we are much better than the Gammas and Deltas. Gammas are stupid.} -- Brave New World, Aldous Huxley
\end{quote}
\begin{quote}
{\it Pessimism of the Intellect, Optimism of the Will} -- Antonio Gramsci
\end{quote}
We have argued that given sufficient phenotype$\vert$genotype data, genomic prediction of traits such as cognitive ability will be possible. If, for example, 0.6 or 0.7 of total population variance is captured by the predictor, the accuracy will be roughly plus or minus half a standard deviation (e.g., a few cm of height, or 8 IQ points). The required sample size to extract a model of this accuracy is probably on the order of a million individuals. As genotyping costs continue to decline, it seems likely that we will reach this threshold within five years for easily acquired phenotypes like height (self-reported height is reasonably accurate), and perhaps within the next decade for more difficult phenotypes such as cognitive ability. At the time of this writing SNP genotyping costs are below \$50 USD per individual, meaning that a single super-wealthy benefactor could independently fund a crash program for less than \$100 million. 

Once predictive models are available, they can be used in reproductive applications, ranging from embryo selection (choosing which IVF zygote to implant) to active genetic editing (e.g., using powerful new CRISPR techniques). In the former case, parents choosing between 10 or so zygotes could improve their expected phenotype value by a population standard deviation. For typical parents, choosing the best out of 10 might mean the difference between a child who struggles in school, versus one who is able to complete a good college degree. Zygote genotyping from single cell extraction is already technically well developed \cite{single}, so the last remaining capability required for embryo selection is complex phenotype prediction. The cost of these procedures would be less than tuition at many private kindergartens, and of course the consequences will extend over a lifetime and beyond.

The corresponding ethical issues are complex and deserve serious attention in what may be a relatively short interval before these capabilities become a reality. Each society will decide for itself where to draw the line on human genetic engineering, but we can expect a diversity of perspectives. Almost certainly, some countries will allow genetic engineering, thereby opening the door for global elites who can afford to travel for access to reproductive technology. As with most technologies, the rich and powerful will be the first beneficiaries. Eventually, though, I believe many countries will not only legalize human genetic engineering, but even make it a (voluntary) part of their national healthcare systems \cite{PGD}. The alternative would be inequality of a kind never before experienced in human history.

Other applications of genomic advances might include the development of drugs which improve cognitive function, or protect against Alzheimer's or Parkinson's disease, or ameliorate conditions such as schizophrenia or autism. From the purely scientific perspective, the elucidation of the genetic architecture of intelligence is a first step towards unlocking the secrets of the brain and, indeed, of what makes humans unique among all life on earth.

%\appendix
%\section{Some title}
%Please always give a title also for appendices.

\acknowledgments

The author thanks Chris Chang, Carson Chow, Chiu-man Ho, Laurent Tellier, Shashaank Vattikuti, and especially James Lee for numerous discussions on the topics covered in these notes. Nevertheless, the author accepts full responsibility for the content. Several of the figures were obtained from James Lee.

%\paragraph{Note added.} This is also a good position for notes added
%after the paper has been written.

\bigskip \bigskip

%\newpage
\noindent  {\bf Frequently Asked Questions} from \url{https://www.cog-genomics.org/faq}

\bigskip
\noindent  {\bf What is g?}

No one knows precisely what intelligence is, and even experts disagree as to how it should be defined. However, it has been known for over a century that performance on different cognitive tests is positively correlated: for example, someone who is good at math puzzles is also more likely to have an above average vocabulary. Given a battery of tests and their correlation matrix, one can use probability theory to define a single parameter that, in a sense, optimally compresses the information from administering them all.

In practice, a wide range of intuitively sensible test batteries and functions of their score vectors yield very similar estimates of this parameter. As a result, psychologists consider these functions of test batteries to all be reasonable estimators of a parameter called the General Factor of Intelligence, or g for short.

\bigskip
\noindent {\bf Why is g important?}

The human brain is extremely complex, a unique product of millions of years of evolution. Our genetic code is the "blueprint" from which this object is constructed. The genetics of cognition inform subjects as diverse as psychology, anthropology, neuroscience and molecular biology. g may be a rough guide to the overall goodness of function of the brain.

In addition, g has significant correlations with health outcomes. It has been demonstrated to be amongst the very best predictors of cardiovascular disease, resistance to dementia, ability to quit smoking, and even simple longevity. The brain is an organ like any other, and it affects human health and well-being profoundly.

\bigskip
\noindent {\bf How precisely can g be defined?}

Note that the definition of g is periodically refined. However, it is undoubtedly a robust phenomenon, and a promising metric upon which to base our study. Some important properties of g are:

{\it Stability}: scores tend not to vary significantly after adolescence,

{\it Heritability}: twin studies and adoption studies powerfully suggest that much of the variance in g is dependent upon genetics,

{\it Predictive power}: g scores are correlated with academic and job performance, income, longevity, etc., even after controlling for other variables such as social class, ethnic background, and resource access.

\bigskip
\noindent {\bf How heritable is g?}

Twin and adoption studies suggest that at least 50\%, and perhaps as much as 80\%, of the variance in g scores is due to genetic causes. Note that heritability is defined over a specific distribution of environments. For the 50-80\% range mentioned above, the environments in question are those found in families typically allowed to adopt foster children. Severe environmental deprivation may reduce heritability, as with most traits, but the predictive value upon g of genetics has long been well established.



\begin{thebibliography}{99}
\bibitem{PGD}
Preimplantation Genetic Diagnosis or Screening (PGD or PGS) of embryos can already be covered by national health care plans in countries such as Israel and Denmark. Predictive modeling of complex traits (including non-Mendelian disease susceptibility) is the likely next advance in PGD/PGS.
\end{thebibliography}
\end{document}